\documentclass[11pt,a4paper]{article}
\pdfoutput=1
\usepackage{jheppub}
\usepackage[T1]{fontenc}
\usepackage{float}
\usepackage{color}
\usepackage{epic}
\begin{document}
\title{Observable Lepton Number Violation with Predominantly Dirac Nature of Active Neutrinos}
\author[a,1]{Debasish Borah \note{Corresponding author}}
\affiliation[a]{Department of Physics, Indian Institute of Technology Guwahati, Assam-781039, India}
\author[b]{Arnab Dasgupta}
\affiliation[b]{Institute of Physics, HBNI, Sachivalaya Marg, Bhubaneshwar-751005, India}

\emailAdd{dborah@iitg.ernet.in}
\emailAdd{arnab.d@iopb.res.in}

\abstract{We study a specific version of $SU(2)_R \times SU(2)_L \times U(1)_{B-L}$ models extended by discrete symmetries where the new physics sector responsible for tiny neutrino masses at leading order remains decoupled from the new physics sector that can give rise to observable signatures of lepton number violation such as neutrinoless double beta decay. More specifically, the dominant contribution to light neutrino masses comes from a one-loop Dirac mass. At higher loop level, a tiny Majorana mass also appears which remains suppressed by many order of magnitudes in comparison to the Dirac mass. Such a model where the active neutrinos are predominantly of Dirac type, also predicts observable charged lepton flavour violation like $\mu \rightarrow 3e, \mu \rightarrow e \gamma$ and multi-component dark matter. }

\maketitle

\section{Introduction}
In spite of significant development in theoretical as well as experimental frontiers of neutrino physics, we still do not know whether neutrinos are of Dirac or Majorana type fermions. The existence of non-zero neutrino masses and their large mixing have been verified again and again at several neutrino oscillation experiments \cite{PDG, kamland08, T2K, chooz, daya, reno, minos} in the last two decades. However, these experiments remain insensitive to the Dirac or Majorana nature of neutrinos. Apart from this, they also can not measure the lightest neutrino mass, leaving open the issue of neutrino mass hierarchy. They can only measure two mass squared differences, three mixing angles and the leptonic Dirac CP violating phase. For the present status of neutrino oscillation parameters, one can refer to the recent global fit analysis in \cite{schwetz14} and \cite{valle14}. The fact that, the standard model (SM) of particle physics can not explain non-zero neutrino masses and mixing, has invited several beyond standard model (BSM) proposals studied extensively in the last few decades.

Since Majorana fermions are their own antiparticles, it will indicate lepton number violation (LNV) in the neutrino sector. This is a typical feature of almost all the BSM proposals put forward to explain non-zero neutrino mass. More popularly known as seesaw mechanisms: type I \cite{ti}. type II \cite{tii0,tii} or type III \cite{tiii}, these frameworks can give rise to tiny neutrino masses of Majorana type by introducing new interactions with LNV through heavy fields. The same heavy fields can also give rise to new sources of lepton flavour violation (LFV) in the charged lepton sector. If the scale of these new particles lies around the TeV corner, the corresponding LNV and LFV contributions should be accessible at the large hadron collider (LHC) searches \cite{lrlhc1, ndbd00}, future collider searches \cite{futureCollider, futureCollider2} as well as rare decay experiments looking for charged lepton flavour violation like $\mu^- \rightarrow e^- e^- e^+, \mu^- \rightarrow e^- \gamma$ \cite{MEG16, sindrum}. Although observing these processes may probe a particular seesaw mechanism responsible for Majorana neutrino masses, the most direct probe of the Majorana nature of light neutrinos is to look for another LNV process called the neutrinoless double beta decay $(0\nu \beta \beta)$ where a heavier nucleus decays into a lighter one and two electrons $(A, Z) \rightarrow (A, Z+2) + 2e^- $ without any (anti) neutrinos in the final state thereby violating lepton number by two units. For a review on $0\nu\beta \beta$, please refer to \cite{NDBDrev}. With the present $0\nu \beta \beta$ experiments like KamLAND-Zen \cite{kamland, kamland2}, GERDA \cite{GERDA, GERDA2} probing the quasi-degenerate regime of light neutrino masses, one can expect the next generation experiments to cover the entire parameter space for $0\nu \beta \beta$, at least in the case inverted hierarchical pattern of light neutrino masses. The current lower limit on the half-life of this rare process from these two experiments lie in the range of $10^{25}-10^{26}$ year. The projected sensitivity of the phase III of KamLAND-Zen is $T_{1/2} > 2\times 10^{26}$ year after two years of data taking. Similar goal is also set by the GERDA experiment to reach $T_{1/2} > 10^{26}$ year. Another experiment called EXO-200 whose 2014 limit was $T_{1/2}> 1.6 \times 10^{25}$ year \cite{exo200} is now anticipating a factor of 2-3 increase in sensitivity after 2-3 years of data taking. Similarly, the next stage of another experiment called CUORE has a projected sensitivity to $T_{1/2} > 9\times 10^{25}$ year. Among the next generation experiments, NEXT-100 has a projected sensitivity of $T_{1/2} > 6 \times 10^{25}$ year whereas Super-NEMO experiment aims to reach sensitivity of $T_{1/2} > 10^{26} $ year. Another experiment called Majorana Demonstrator will reach similar sensitivity in three years. Similarly, AMoRe experiment is expected to achieve a sensitivity of $T_{1/2} > 3\times 10^{26} $ year. A comprehensive summary of these ongoing and upcoming experimental efforts can be found in the recent article \cite{0nbbexpt}.

The absence of any positive signal at $0\nu \beta \beta$ experiments does not necessarily rule out the Majorana nature of light neutrinos. For example, the light neutrino contributions to $0\nu \beta \beta$ can remain very much suppressed for certain range of parameters if neutrinos obey a normal hierarchical pattern. The contribution can even be zero, when the $ee$ element of the Majorana neutrino mass matrix vanishes (To know more about the possible zeros in light neutrino mass matrix, please refer to \cite{Ludl2014}). On the other hand, a positive signal at $0\nu \beta \beta$ guarantees a non-zero effective Majorana mass for the electron type neutrino according to the Schechter-Valle theorem \cite{schvalle}. Although one can introduce some cancellations between different terms leading to a vanishing effective Majorana mass, one can not guarantee such cancellations to all orders of perturbation theory. In fact, there exists no continuous or discrete symmetry that can forbid such an effective Majorana mass term to all orders in perturbation theory \cite{takasugi}. The quantitative impact of the Schechter-Valle theorem was investigated by the authors of \cite{4loopcomp} and found that the maximum contribution to effective Majorana mass of electron type neutrino from a non-zero $0\nu \beta \beta$ amplitude is of the order of $10^{-28}$ eV, way below the scale at which light neutrino masses lie. This leads to a very important conclusion that the new physics sector responsible for LNV processes like $0\nu \beta \beta$ may not be related to the new physics sector responsible for leading order contribution to light neutrino masses. Although an example of such a scenario appeared in \cite{GB1}, we do not see much work in particle physics literature pursuing such a possibility. Motivated by this, here we propose a model where the new physics sector can give rise to observable $0\nu \beta \beta$ and LNV signatures at colliders although the light neutrino mass remains predominantly of Dirac type with a negligible Majorana type contribution. The model also predicts observable charged lepton flavour violation, multi-component dark matter and matter-antimatter asymmetry of the Universe. We constrain the parameter space of the model from the requirement of satisfying correct neutrino and dark matter data and also predict new signatures at $0 \nu \beta \beta$ and LFV experiments.

This paper is organised as follows. In section \ref{model}, we discuss our model followed by a discussion on the generation of tiny neutrino mass at one-loop level in section \ref{numass}. In section \ref{sec3}, we discuss possible new physics contribution to neutrinoless double beta decay and then discuss charged lepton flavour violation in section \ref{sec4}. We discuss about the possible dark matter candidates and the standard calculation of dark matter relic abundance in section \ref{sec5}. We briefly comment on the possibility of active-sterile oscillations over astronomical distances due to tiny pseudo-Dirac splittings in section \ref{sec06} and finally discuss our results in section \ref{sec6}.

\section{The Model}
\label{model}
The model we propose in this work is an extension of the popularly known left-right symmetric models (LRSM) \cite{lrsm, lrsmpot} studied extensively in the literature. In these models, the gauge symmetry of the electroweak theory is extended to $SU(3)_c \times SU(2)_L \times SU(2)_R \times U(1)_{B-L}$. The right handed fermions are doublets under $SU(2)_R$ similar to the way left handed fermions transform as doublets under $SU(2)_L$. The requirement of an anomaly free $U(1)_{B-L}$ makes the presence of right handed neutrinos a necessity rather than a choice. Since the minimal version of this model predicts Majorana nature of light neutrinos by virtue of the in built seesaw mechanism, we consider a version of LRSM where the tree level Majorana mass term for the light neutrinos can be forbidden. One such possibility lies in the LRSM without the conventional Higgs bidoublet \cite{VLQlr, univSeesawLR} where all the fermions acquire masses through a universal seesaw mechanism due to the presence of additional heavy fermions. Very recently this model was also studied in the context of 750 GeV di-photon excess at LHC \cite{lhcrun2a,atlasconf,CMS:2015dxe} \footnote{It should be noted that the latest updates from the LHC experiments \cite{LHC16update} do not confirm their preliminary hints towards this 750 GeV di-photon resonance.} by several authors \cite{LR750GeV1, LR750GeV2, LR750GeV3, LR750GeV4}. As shown recently \cite{db16}, the heavy fermions introduced to generate light neutrino masses can have some non-trivial transformations under additional discrete symmetries such that, a tiny Dirac neutrino mass can be generated at one-loop level through \textit{scotogenic} fashion \cite{m06}. The scalar fields of $SU(2)_L$ and $SU(2)_R$ sectors do not necessarily have the same transformations under the additional discrete symmetries thereby deviating from the purely left-right symmetric limit of the conventional LRSM.
\begin{table}
\begin{center}
\begin{tabular}{|c|c|c|}
\hline
Particles & $SU(3)_c \times SU(2)_L \times SU(2)_R \times U(1)_{B-L} $  & $Z_4 \times Z_4$  \\
\hline
$q_L=\begin{pmatrix}u_{L}\\
d_{L}\end{pmatrix}$ & $(3, 2, 1, \frac{1}{3})$ & $(1, 1)$ \\
$q_R=\begin{pmatrix}u_{R}\\
d_{R}\end{pmatrix}$ & $(3, 1, 2, \frac{1}{3})$ & $(1, 1)$ \\
$\ell_L=\begin{pmatrix}\nu_{L}\\
e_{L}\end{pmatrix}$ & $(1, 2, 1, -1)$ & $(1, 1)$ \\
$\ell_R=\begin{pmatrix}N_{R}\\
e_{R}\end{pmatrix}$ & $(1, 2, 1, -1)$ & $(1, 1)$ \\
$U_{L,R}$ & $(3, 1, 1, \frac{4}{3})$ & $(1,1)$ \\
$D_{L,R}$ & $(3, 1, 1, -\frac{2}{3})$ & $(1,1)$ \\
$ E_{L,R}$ & $(1,1,1, -2)$ & $(1,1)$ \\
$\nu_R$ & $(1,1,1,0)$ & $(1,i)$ \\
$ \psi_{L,R}$ & $(1,1,1, 0)$ & $(i,1)$ \\
\hline
\end{tabular}
\end{center}
\caption{Fermion Content of the Model}
\label{tab:data1}
\end{table}
\begin{table}
\begin{center}
\begin{tabular}{|c|c|c|}
\hline
Particles & $SU(3)_c \times SU(2)_L \times SU(2)_R \times U(1)_{B-L} $  & $Z_4 \times Z_4$  \\
\hline
$H_L=\begin{pmatrix}H^+_{L}\\
H^0_{L}\end{pmatrix}$ & $(1,2,1,-1)$ & $(1,1)$  \\
$H_R=\begin{pmatrix}H^+_{R}\\
H^0_{R}\end{pmatrix}$ & $(1,1,2,-1)$ & $(1,1)$ \\
$\eta_L=\begin{pmatrix}\eta^+_{L}\\
\eta^0_{L}\end{pmatrix}$ & $(1,2,1,-1)$ & $(-i,1)$  \\
$\eta_R=\begin{pmatrix}\eta^+_{R}\\
\eta^0_{R}\end{pmatrix}$ & $(1,1,2,-1)$ & $(-i,-1)$ \\
$\Delta_R=\begin{pmatrix} \delta_{R}^+/\sqrt{2} & \delta_{R}^{++} \\ \delta_{R}^0 & -\delta_{R}^+/\sqrt{2} \end{pmatrix}$ & $(1,1,3,2)$ & $(1,1)$ \\
$\Delta_L=\begin{pmatrix} \delta_{L}^+/\sqrt{2} & \delta_{L}^{++} \\ \delta_{L}^0 & -\delta_{L}^+/\sqrt{2} \end{pmatrix}$ & $(1,3,1,2)$ & $(1,-1)$ \\
$ \chi_1 $ & $(1,1,1,0)$ & $(-i, i)$\\
$ \chi_2 $ & $(1,1,1,0)$ & $(1, i)$\\
$ \chi_3 $ & $(1,1,1,0)$ & $(-1, -1)$\\
\hline
\end{tabular}
\end{center}
\caption{Scalar content of the Model}
\label{tab:data2}
\end{table}

The particle content of the model is shown in table \ref{tab:data1} and \ref{tab:data2}. In the fermion content shown in table \ref{tab:data1}, the doublets are the usual LRSM fermion doublets and the vector like fermions $U, D, E$ are required for the universal seesaw for charged fermion masses. The gauge singlet fermions $\nu_R, \psi$ are chosen to generate neutrino masses at one loop order, similar to the way it was shown in \cite{db16} within LRSM and more recently in \cite{dbad1}. Their transformations under the additional discrete symmetry $Z_4 \times Z_4$ are chosen in such a way that their Majorana mass terms are forbidden. Among the scalar fields, shown in table \ref{tab:data2}, $H_{L,R}$ are needed to break the gauge symmetry all the way down to the $SU(3)_c \times U(1)_Q$ leading to heavy vector bosons $W_{L,R}, Z_{L,R}$. The scalar $\Delta_{R}$ imparts Majorana mass term to the neutral fermion of the right handed lepton doublets whereas $\Delta_L$ does not couple to the leptons due to the chosen discrete charges. Both of these scalar triplets however, contribute to the vector boson masses. The additional scalar doublets $\eta_{L,R}$ are there to provide the dark matter candidates as well as neutrino mass because the left handed doublet $\eta_L$ goes inside the one-loop diagram for Dirac neutrino mass as we discuss below. The discrete charges of $\eta_R$ are chosen in a way that prevents similar one-loop Dirac neutrino mass diagram between $N_R$ and $\nu_R$. This is done in order to keep the major source of LNV (In our model $\Delta_R$ and $N_R$) decoupled from the source of neutrino mass at leading order. The two of the three singlet scalars namely, $\chi_{1,2}$ are needed to complete the one-loop neutrino mass diagram. Although, as such the presence of $\Delta_L, \eta_R, \chi_3$ may look redundant, they have non-trivial role to play in dark matter phenomenology as we discuss later.

The Lagrangian for fermions can be written as
\begin{align}
\mathcal{L} & \supset Y_U (\overline{q_L} H_L U_L+\overline{q_R} H_R U_R) + Y_D (\overline{q_L} H^{\dagger}_L D_L+\overline{q_R} H^{\dagger}_R D_R) +M_U \overline{U_L} U_R+ M_D \overline{D_L} D_R\nonumber \\
& +Y_E (\overline{\ell_L} H^{\dagger}_L E_L+\overline{\ell_R} H^{\dagger}_R E_R) +M_E \overline{E_L} E_R+Y_\nu \overline{\ell_L} \eta_L \psi_R+ M_\psi \overline{\psi_L} \psi_R + Y_r \overline{\nu_R} \chi_1 \psi_L \nonumber \\
& +f_{R} \ell_{R}^T \ C \ i \sigma_2 \Delta_R \ell_{R}+\text{h.c.}
\label{fermionL}
\end{align}
The relevant part of the scalar Lagrangian is
\begin{align}
\mathcal{L} & \supset -\mu^2_L H^{\dagger}_L H_L + \lambda_L (H^{\dagger}_L H_L)^2 -\mu^2_R H^{\dagger}_R H_R + \lambda_R (H^{\dagger}_R H_R)^2 +\mu^2_{\eta_L} \eta^{\dagger}_L \eta_L + \lambda_{\eta_L} (\eta^{\dagger}_L \eta_L)^2 \nonumber \\
& +\mu^2_{\eta_R} \eta^{\dagger}_R \eta_R + \lambda_{\eta_R} (\eta^{\dagger}_R \eta_R)^2-\mu^2_{\Delta_L} \Delta^{\dagger}_L \Delta_L + \lambda_{\Delta_L} (\Delta^{\dagger}_L\Delta_L)^2-\mu^2_{\Delta_R} \Delta^{\dagger}_R \Delta_R + \lambda_{\Delta_R} (\Delta^{\dagger}_R\Delta_R)^2  \nonumber \\
& + \mu^2_1 \chi^{\dagger}_1 \chi_1 + \lambda_1 ( \chi^{\dagger}_1 \chi_1)^2-\mu^2_2 \chi^{\dagger}_2 \chi_2 + \lambda_2 ( \chi^{\dagger}_2 \chi_2)^2+ \mu_3 H_R H_R \Delta_R + \lambda_3 \eta^{\dagger}_L H_L \chi_1 \chi^{\dagger}_2 \nonumber \\
& + \lambda_4 \eta_L \eta_L \Delta_L \chi_3 + \mu_4 \chi_1 \chi_1 \chi_3 + \lambda_{5L,R} (H^{\dagger i}_{L,R}H_{L,R i})(\eta^{\dagger j}_{L,R}\eta_{L,R j})+\lambda_{6 L,R} 
(H^{\dagger i}_{L,R} H_{L,R j})(\eta^{\dagger j}_{L,R}\eta_{L,R i})
\label{scalarL}
\end{align} 

We denote the vacuum expectation value (vev) acquired by the neutral components of the fields responsible for spontaneous gauge symmetry breaking as $\langle H^0_L \rangle = v_L/\sqrt{2}, \langle H^0_R \rangle = v_R/\sqrt{2}, \langle \delta^0_L \rangle = v_{\delta_L}/\sqrt{2}, \langle \delta^0_R \rangle = v_{\delta_R}/\sqrt{2}$. The gauge symmetry breaking is achieved as 
$$SU(2)_L \times SU(2)_R \times U(1)_{B-L} \quad \underrightarrow{\langle
H_R, \Delta_R \rangle} \quad SU(2)_L\times U(1)_Y  \quad \underrightarrow{\langle H_L \rangle} U(1)_{Q}$$
Here we have omitted $SU(3)_c$ which remains unbroken throughout the above symmetry breaking stages. After this symmetry breaking, the electromagnetic charge of the components of above fields arise as
\begin{align}
Q=T_{3L}+T_{3R}+\frac{B-L}{2} 
\end{align}
These charges are shown as superscripts of different scalar fields in table \ref{tab:data2}. As a result of this symmetry breaking, two charged and two neutral vector bosons acquire masses. The mass matrix squared for charged gauge bosons in the basis $W^{\pm}_L, W^{\pm}_R$ is 
\begin{equation}
M^2_{\pm} = \frac{1}{4}\begin{pmatrix} g^2_L (v^2_L+2v^2_{\delta_L}) & 0 \\ 0 & g^2_R (v^2_R+2v^2_{\delta_R}) \end{pmatrix}
\end{equation}
Similarly, the neutral gauge boson mass matrix in the basis $(W_{L3}, W_{R3}, B)$ is
\begin{equation}
M^2_{0} = \frac{1}{4}\begin{pmatrix} g^2_L (v^2_L+4v^2_{\delta_L}) & 0 & -g_1 g_L (v^2_L+4v^2_{\delta_L}) \\ 0 & g^2_R (v^2_R+4v^2_{\delta_R}) & -g_1 g_R (v^2_R+4v^2_{\delta_R})\\  -g_1 g_L (v^2_L+4v^2_{\delta_L}) & -g_1 g_R (v^2_R+4v^2_{\delta_R}) & g^2_1(v^2_L+v^2_R+4v^2_{\delta_L}+4v^2_{\delta_R})\end{pmatrix}
\end{equation}
Here we have denoted the gauge couplings of $SU(2)_L, SU(2)_R, U(1)_{B-L}$ gauge groups as $g_L, g_R, g_1$. In the left-right symmetric limit, $g_L = g_R$. Assuming $v_{\delta_L} \ll v_L \ll v_R, v_{\delta_R}$ and $g_L=g_R=g$, we can write down the vector boson masses as
$$ M_{W_L} \approx \frac{gv_L}{2}, \;\;\; M_{W_R} =\frac{g}{2}\sqrt{v^2_R + 4v^2_{\delta_R}} $$
$$M_{Z_L} \approx \frac{g v_L}{2} \sqrt{1+ \frac{g^2_1}{g^2+g^2_1}}, \;\;\; M_{Z_R} \approx \frac{1}{2} \sqrt{(g^2+g^2_1)(v^2_R + 4 v^2_{\delta_R})} $$
Since there exists no scalar fields simultaneously charged under $SU(2)_L$ and $SU(2)_R$ (like the bidoublet scalar in minimal LRSM), here we do not have any tree level $W_L-W_R$ mixing. It should be noted that, the equality of gauge couplings $g_L = g_R$ is no longer guaranteed by the in built symmetry of the model. However, we consider it as a benchmark point so as to apply the conservative lower bounds on the masses of heavy gauge bosons and scalar particles of the model from the LHC experiment, to be discussed below. Also, the smallness of the vev of the neutral component of $\Delta_L$ does not arise naturally in the form of an induced vev after electroweak symmetry breaking. This is due to the absence of trilinear coupling of the form $H_L H_L \Delta_L$ in the model. However, one needs to keep the vev of left triplet scalar small as the constraints from electroweak $\rho$ parameter restricts it to $v_{\delta_L} \leq 2$ GeV \cite{pdg15}. In the Standard Model, the $\rho$ parameter is unity at tree level, given by 
$$ \rho = \frac{M^2_{W_L}}{M^2_{Z_L} \cos^2 \theta_W} $$
where $\theta_W$ is the Weinberg angle. But in the presence of left scalar triplet vev, there arises additional contribution to the electroweak gauge boson masses which results in a departure of the $\rho$ parameter from unity at tree level.
$$ \rho = \frac{1+\frac{2v^2_{\delta_L}}{v^2_L}}{1+\frac{4v^2_{\delta_L}}{v^2_L}} $$
Experimental constraints on the $\rho$ parameter $\rho = 1.00040 \pm 0.00024$ \cite{pdg15} forces one to have $v_{\delta_L} \leq 2$ GeV. Since, this can not be generated as an induced vev (which can be naturally small), one has to fine tune the quartic couplings and bare mass term of $\Delta_L$ scalar in order to generate such a small vev.

The charged fermion masses appear after integrating out the heavy vector like charged fermions. After integrating out the heavy fermions, the charged fermions of the standard model develop Yukawa couplings to the scalar doublet $H_L$ as follows
$$ y_u = Y_U \frac{v_R}{M_U} Y^T_U, \;\;y_d = Y_D \frac{v_R}{M_D} Y^T_D, \;\;y_e = Y_E \frac{v_R}{M_E} Y^T_E $$
The apparent seesaw then can explain the observed mass hierarchies among the three generations of charged fermions. The vector-like fermion masses appearing in the above relations are however, tightly constrained from direct searches. For example, the vector like quark masses have a lower limit $m_q \geq 750-920$ GeV depending on the particular channel of decay \cite{VLQconstraint} whereas this bound gets relaxed to $m_q \geq 400$ GeV \cite{VLQconstraint2} for long lived vector like quarks. These exclusion ranges slightly get changed in the more recent LHC exclusion results on vector like quarks: $m_q  > 810-1090$ GeV where the vector like quarks decaying into W bosons and b quarks n the lepton plus jet final state was searched for at 13 TeV centre of mass energy \cite{VLQconstraint3}. Another 13 TeV search for vector like top quarks using final states of one lepton, at least four jets and large missing transverse momentum puts limit on vector like top partner masses as $m_q > 810-1130$ GeV \cite{VLQconstraint4}. Further constraints on vector like quarks can be found in \cite{vlqhandbook}. The constraints on vector like leptons are much weaker $m_l \geq 114-176$ GeV \cite{VLLconstraint}. These vector like fermions also get constrained from electroweak precision data by virtue of their contributions to the oblique correction parameters $S, T, U$ \cite{Lavoura&Silva}. The experimental bound on these oblique parameters \cite{pdg15} can be satisfied if we consider a conservative upper bound on the mixing of vector like fermions with the SM fermions as $\sin \theta \lesssim 0.1$. For the quarks, this will imply 
\begin{equation}
\sin \theta = \sqrt{ \frac{m_q v_R}{v_L M}} \lesssim 0.1.
\label{ewpt}
\end{equation}
where we have considered  that $\theta$ is the mixing between the SM quark $q$ with mass $m_q$ and the corresponding heavy vector like quark with mass $M$. In the minimal model with only $H_{L, R}$ as scalars, we have $v_L \approx 246 $ GeV and $v_R \geq 6$ TeV, for $M_{W_R} \geq 3$ TeV. Now, for the bottom quark as an example, this bound will imply the corresponding vector like quark mass to be heavier than 10 TeV. Since we have two separate scalar fields contributing to the right handed gauge boson masses with only one of them contributing to the charged fermion masses, we can tune $v_R$ to a lower value while keeping $v_{\delta_R} \approx 6$ TeV for a 3 TeV $W_R$ boson. This will enable us to satisfy the above bound  \eqref{ewpt} without taking the vector like fermion masses beyond the TeV scale. The neutral fermion $N_R$ which is a part of the right handed lepton doublet $\ell_R$ acquires a Majorana mass term $M_R = f_R v_{\delta_R}$. The active neutrinos $\nu_L$ which are part of left handed lepton doublets $\ell_L$ remain massless along with singlet neutrinos $\nu_R$ at tree level. However, they acquire a Dirac mass at one loop level as shown in figure \ref{numass} to be discussed in the next section.

Apart from the vector like fermions, the experimental constraints on other particles in the model, particularly the right handed gauge bosons, triplet scalar and neutral fermion from right handed lepton doublets should also be taken into account. The right handed gauge boson masses are primarily constrained from $K-\bar{K}$ mixing and direct searches at the LHC. While $K-\bar{K}$ mixing puts a constraint $M_{W_R} > 2.5$ TeV \cite{kkbar}, direct search bounds depend on the particular channel under study. For example, the dijet resonance search in ATLAS experiment puts a bound $M_{W_R} > 2.45$ TeV at $95\%$ CL \cite{dijetATLAS} in the $g_L=g_R$ limit. On the other hand, the CMS search for same sign dilepton plus dijet $pp \rightarrow l^{\pm} l^{\pm} j j$ mediated by heavy right handed neutrinos at 8 TeV centre of mass energy excludes some parameter space in the $M^{\text{lightest}}_i-M_{W_R}$ plane \cite{CMSNRWR} where $M^{\text{lightest}}_i$ is the the mass of the lightest neutral fermion from right handed lepton doublets. More recently, the results on dijet searches at ATLAS experiment at 13 TeV centre of mass energy has excluded heavy W boson masses below 2.9 TeV \cite{dijetATLAS2}. Similarly, the doubly charged scalar (from left scalar triplet) also faces limits from CMS and ATLAS experiments at LHC:
$$ M_{\Delta^{\pm \pm}} \geq 445 \; \text{GeV} \; (409 \; \text{GeV}) \; \text{for} \; \text{CMS (ATLAS)} $$
These limits have been put by assuming $100\%$ leptonic branching factions \cite{hdlhc}. The limits on doubly charged scalars have been updated recently from 13 TeV data as: $ M_{\Delta^{\pm \pm}_L} \geq 570 \; \text{GeV}, M_{\Delta^{\pm \pm}_R} \geq 420 \; \text{GeV}$ \cite{hdlhc2} assuming $100\%$ branching ratio into electrons. For $50\%$ branching ration into electrons, these limits get slightly relaxed $ M_{\Delta^{\pm \pm}_L} \geq 530 \; \text{GeV}, M_{\Delta^{\pm \pm}_R} \geq 380 \; \text{GeV}$ \cite{hdlhc2}. These limits will be relaxed further for lower leptonic branching ratios, like in the present model, where the left handed doubly charged scalar has no tree level couplings to the leptons.

There also exists bounds from $0\nu \beta \beta$ and LFV decay processes $\mu \rightarrow 3e, \mu \rightarrow e \gamma$ on the masses of heavy neutral fermions $M_i$ as well as triplet scalar masses $M_{\Delta}$. Earlier, it was shown \cite{ndbd00} that existing experimental bounds on these decay processes forces triplet masses to be at least ten times heavier than the heaviest neutral fermion mass $M_i/M_{\Delta} < 0.1$ if the neutrino mass is generated from either type I or type II seesaw. A more recent work \cite{ndbd102} showed the possibility of lighter triplet scalars $M_i/M_{\Delta} \approx 0.5$. In a subsequent work \cite{ndbd103}, it was shown that one can also have the possibility of $M_i/M_{\Delta} > 1$ if we consider the new physics contribution to the above-mentioned decay processes within a framework of equally dominant type I and type II seesaw, earlier studied in this context by \cite{ndbd101}. Due to a different way of generating leading order neutrino mass in the present model, these bounds may however change as we discuss in the upcoming sections with further details.
\begin{figure}[htb]
\centering
\includegraphics[scale=0.75]{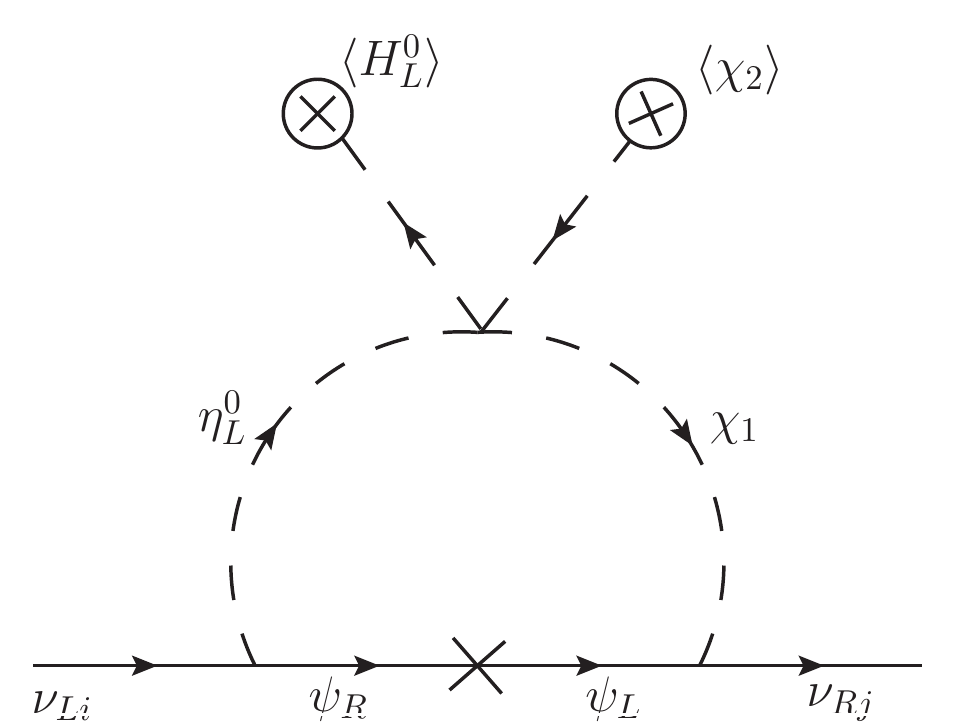}
\caption{One-loop contribution to Dirac neutrino mass}
\label{numass}
\end{figure}
\section{Neutrino Masses}
\label{numass}
The dominant contribution to active neutrino mass comes from the one-loop diagram shown in figure \ref{numass}. Similar one loop diagram for Dirac mass was also discussed in \cite{ma1, db16, dbad1}. Following the one loop computation shown in \cite{ma1, dbad1}, the light neutrino mass can be written as
\begin{equation}
(m_{\nu})_{ij} = (m_{\nu})_{Rij}+(m_{\nu})_{Iij}
\end{equation}
where the two terms on the right hand side with subscript $R, I$ correspond to the contribution from real and imaginary parts of the internal scalar fields respectively. The complex scalar fields in the internal lines can be written in terms of their real and imaginary parts as $\eta^0_L = (\text{Re}(\eta^0_{L}) +i \text{Im}(\eta^0_{L}))/\sqrt{2}, \chi_1 = (\text{Re}(\chi_1) + i \text{Im}(\chi_1))/\sqrt{2}$. The contribution of the real sector $\text{Re}(\eta^0_{L}), \text{Re}(\chi_1)$ to one loop Dirac neutrino mass can be written as 
\begin{equation}
(m_{\nu})_{Rij} =\frac{\sin{\theta_1} \cos{\theta_1}}{32 \pi^2} \sum_k (Y_{\nu})_{ik}(Y_r)_{kj}M_{\psi k} \left ( \frac{m^2_{\xi_1}}{m^2_{\xi_1}-M^2_{\psi k}} \text{ln} \frac{m^2_{\xi_1}}{M^2_{\psi k}}-\frac{m^2_{\xi_2}}{m^2_{\xi_2}-M^2_{\psi k}} \text{ln} \frac{m^2_{\xi_2}}{M^2_{\psi k}} \right)
\label{numassMR}
\end{equation}
where $\xi_{1,2}$ denote the physical mass eigenstates of the $\text{Re}(\eta^0_{L}), \text{Re}(\chi_1)$ sector with a mixing angle $\theta_1$. This mixing angle is related to the mass terms of the scalar potential as well as to the quartic coupling $\lambda_3 \eta^{\dagger}_L H_L \chi_1 \chi^{\dagger}_2 $ involved in the one loop diagram shown in figure \ref{numass} as
$$\tan{2 \theta_1} = \frac{\lambda_3 v_L u}{m^2_{\text{Re}(\chi_1)}-m^2_{\text{Re}(\eta^0_{L})}} $$
Here $v_L/\sqrt{2}, u/\sqrt{2}$ are the vev's of $H^0_L, \chi_2$ respectively. Similar expressions can be written for the contribution of imaginary components of the internal scalar fields to the neutrino mass, as discussed in the recent work \cite{dbad1}. Considering the new physics sector to lie around the TeV scale or equivalently for example, $m_{\xi_1} = 100$ GeV and $M_{\psi} = 10$ TeV, the first term on the right hand side of the equation \eqref{numassMR} becomes 
$$ (m_{\nu})^{1}_{Rij} = 1.46 \times 10^{-2} \sin{2 \theta_1} \sum_k (Y_{\nu})_{ik}(Y_r)_{kj} \; \text{GeV}$$
which can remain at the sub-eV scale if 
\begin{equation}
\sin{2 \theta_1} (Y_{\nu})_{ik}(Y_r)_{kj}  < 10^{-8} 
\label{cond1}
\end{equation}
Which can be easily satisfied by suitable choice of Yukawa couplings as well as quartic coupling generating the mixing angle $\theta_1$.

\begin{figure}[htb]
\centering
\includegraphics[scale=0.95]{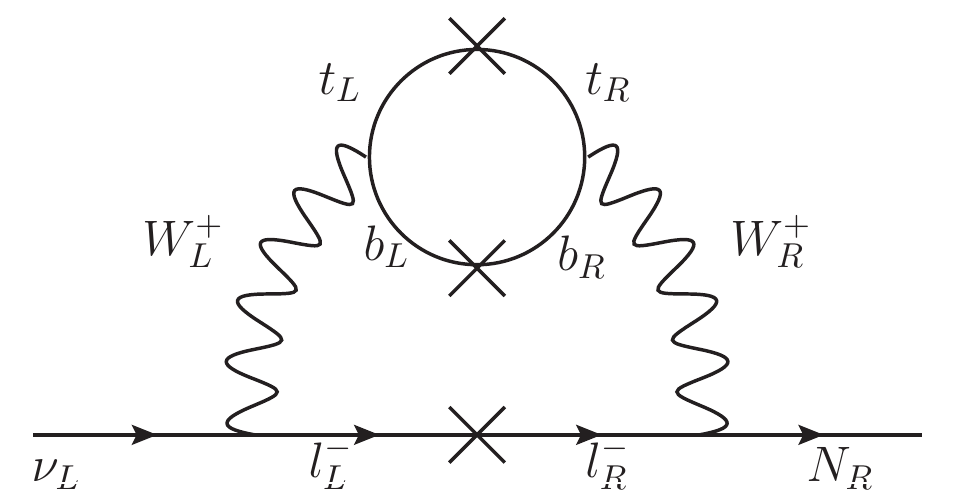}
\caption{Two-loop contribution to Dirac neutrino mass}
\label{numass2}
\end{figure}

The active neutrinos, which are part of the left handed lepton doublets $\ell_L$, acquire a non-zero Dirac mass through its mixing with singlet neutrinos $\nu_R$ at one loop level, as discussed above. The neutral fermions $N_R$, part of the right handed lepton doublets $\ell_R$ acquire non-zero Majorana masses through the vev of the the neutral component of scalar triplet $\Delta_R$. The choice of discrete symmetries prevents the generation of a tree level Majorana mass term of the active neutrinos, due to the absence of $\ell_L - \Delta_L$ couplings. Similarly the choice of singlet scalars in the model, does not give rise to Majorana mass terms of the left and right handed components of the vector like fermions $\psi$. On the other hand, the neutral fermion $N_R$ does not mix with $\nu_R$ at one loop level like the way $\nu_L$ and $\nu_R$ mixes at one loop level. Therefore, upto one loop order, the active neutrinos $\nu_L$ acquire a tiny Dirac mass only through its mixing with $\nu_R$. However, $\nu_L$ can acquire a Dirac mass through mixing with $N_R$ at two loop level, as seen from figure \ref{numass2}. The contribution of this diagram was first computed by \cite{babuhe} \footnote{Here we note that a more realistic possibility of Dirac neutrino mass through such $W_L-W_R$ mixing diagrams was considered very recently by the authors of \cite{ma2016}.} and was found to be approximately
\begin{equation}
M_{LR} \approx \frac{\alpha m_{l^-}}{4 \pi \sin^2{\theta_W}} \theta_{L-R} I
\end{equation}
where $I$ is the loop integration factor (of the order $1-10$) and $\theta_{L-R}$ is the one loop mixing between $W_L, W_R$ given by 
\begin{equation}
\theta_{L-R} \approx \frac{\alpha}{4 \pi \sin^2{\theta_W}} \frac{m_b m_t}{M^2_{W_R}}
\end{equation}
Using $\alpha = 1/137, \sin^2{\theta_W} \approx 0.23, m_b \approx 4.2\; \text{GeV}, m_t \approx 174 \; \text{GeV}, M_{W_R} \approx 3\; \text{TeV}$, we find $\theta_{L-R} \approx 2 \times 10^{-7}$. Using this in the expression for Dirac mass we get 
\begin{equation}
M_{LR} \approx (1-10) \times 5.2 \times 10^{-10} m_{l^-}
\end{equation}
which, for $m_{l^-} = m_e \approx 0.5 \; \text{MeV}$ becomes $M_{LR} \approx (1-10) \times 2.6 \times 10^{-4} \; \text{eV}$. On the other hand, for $m_{l^-} = m_{\tau} \approx 1.77 \; \text{GeV}$, the Dirac mass becomes $M_{LR} \approx (1-10) \times 0.92 \; \text{eV}$. Such a Dirac mass term generates a type I seesaw mass matrix in the $(\nu_L, N_R)$ basis, given by
\begin{equation}
\mathcal{M}_\nu= \left( \begin{array}{cc}
              0 & M_{LR}   \\
              M^T_{LR} & M_{RR}
                      \end{array} \right) \, ,
\label{eqn:numatrix}       
\end{equation}
Using the approximation $M_{RR} \gg M_{LR}$, the light neutrino mass is given by the type I seesaw formula 
\begin{equation}
M_{\nu}^I=-M_{LR} M_{RR}^{-1} M_{LR}^{T}
\label{type1eq}
\end{equation}
where $M_{RR} = f_R v_{\delta_R}$ is the Majorana mass matrix of $N_R$. In this model $M_{LR} < 1 \; \text{eV}$ as discussed above. Therefore, even if we consider a minimal mass of 1 GeV for $N_R$, the corresponding Majorana mass term for active neutrinos is of the order of $10^{-9}$ eV, around eight order of magnitudes suppressed compared to the expected mass of around $0.1$ eV. Although we have used the approximate formula for this two loop Dirac mass from \cite{babuhe} for qualitative understanding, we derive the exact formula for numerical analysis. This is given by 
\begin{align}
M_{LR} &= \frac{\alpha m_{l^-}}{4 \pi \sin^2{\theta_W}} \frac{\sin{2 \theta_{L-R}}}{2} \left(f(x_{l,W_R}) - f(x_{l,WL})\right) \\
\sin{2\theta_{L-R}} &=\frac{2W_{LR}}{\sqrt{\left(M^2_{W_R}-M^2_{W_L}\right)^2 + 4W^2_{LR}}} \nonumber \\ 
W_{LR} &= \frac{4\pi \alpha}{\sin^2 \theta_{W}}\sum_{u,d}m_u m_d V_{u,d}V^*_{u,d}f(x_{u,d}); \quad x_{i,j} = \frac{m^2_i}{m^2_j}\nonumber \\
f(x_{i,j}) &= \frac{1}{16\pi^2}\left[\frac{x_{i,j}\ln (x_{i,j}) + 1 - x_{i,j}}{1-x_{i,j}}+\ln \left(\frac{\mu^2}{m^2_{j}}\right)\right] \nonumber
\end{align}
Therefore, the active neutrino masses are dominantly of Dirac type with tiny signature of lepton number violation. However, there can be observable signatures of lepton number violation through neutrinoless double beta decay as will be discussed below; but the contribution of such lepton number violating physics to Majorana mass of active neutrinos remain suppressed.
\begin{figure}[htb]
\centering
\includegraphics[scale=0.75]{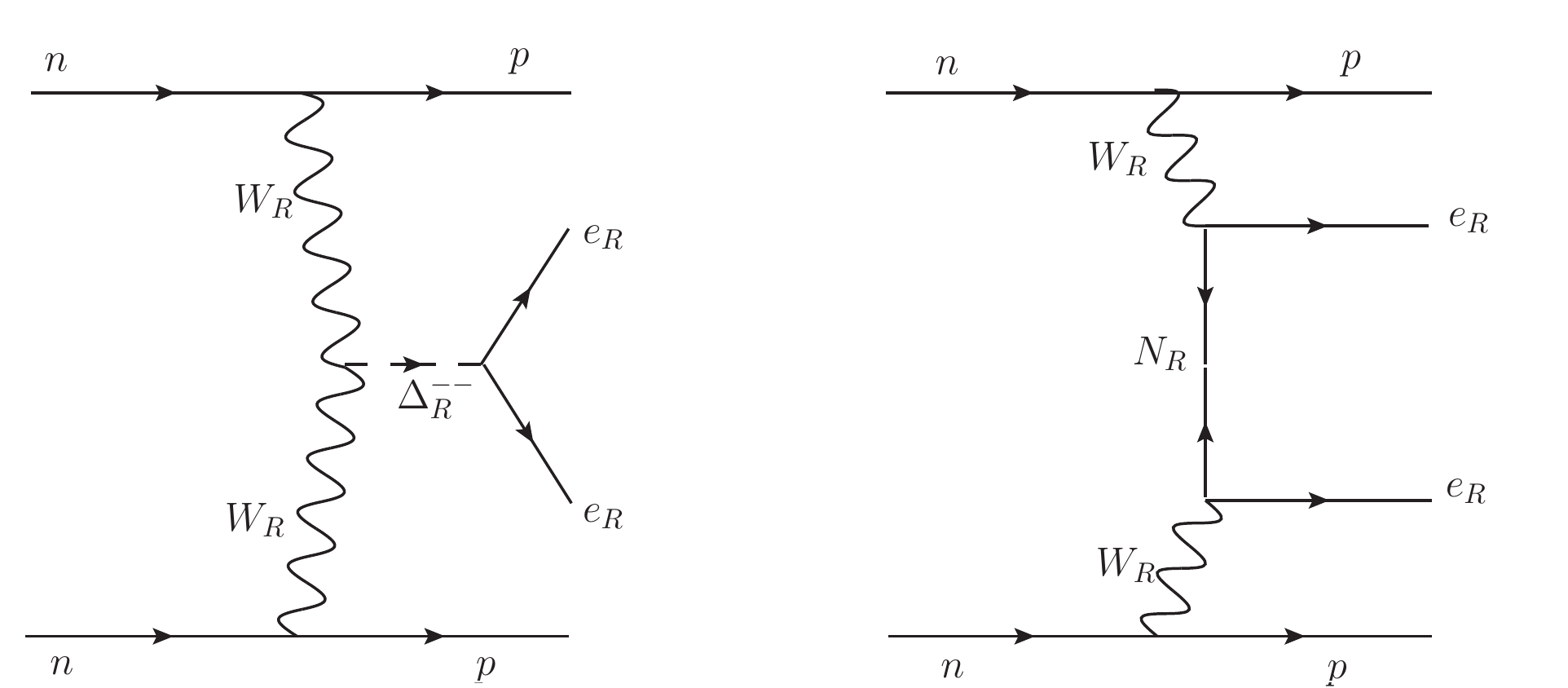}
\caption{Leading Contribution to Neutrinoless Double Beta Decay}
\label{ndbd1}
\end{figure}

\begin{figure}[htb]
\centering
\includegraphics[scale=1.0]{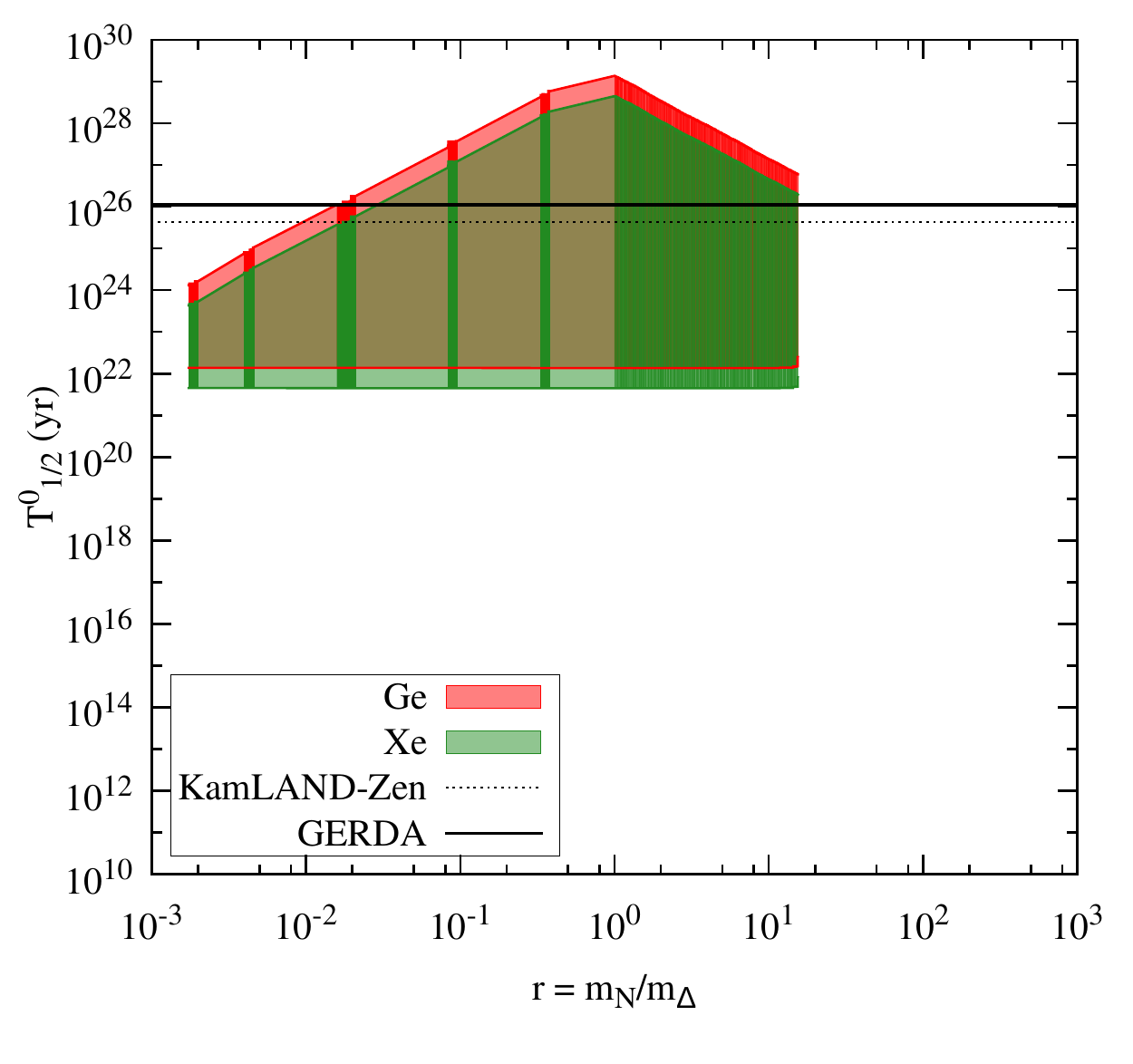}
\caption{Half-life of $0\nu \beta \beta$ as a function of $r=\frac{M_N}{M_{\Delta}}$, the ratio of the masses of heavy neutrino and that of the doubly charged scalar from the triplet $\Delta_R$. The chosen parameters are $M_{W_R} = 3$ TeV, $M_N \in 1-6000$ GeV, $M_{\Delta^{\pm \pm}_R} \in 420-6000$ GeV.}
\label{Thalf}
\end{figure}

\begin{figure}[htb]
\centering
\includegraphics[scale=0.7]{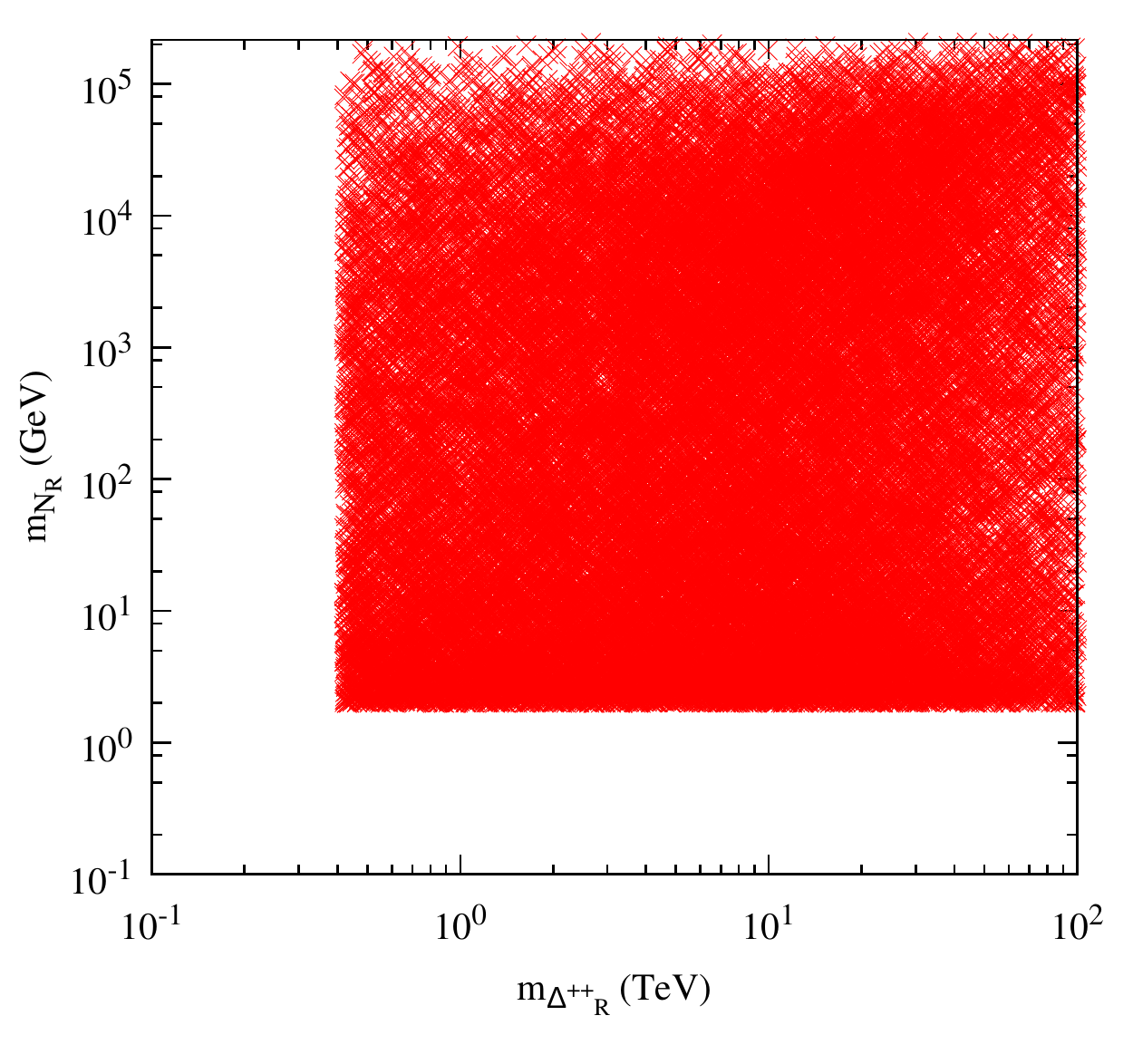}
\caption{Allowed Parameter space in heavy neutrino versus $\Delta^{\pm \pm}_R$ mass from KamLAND-Zen bound on $0\nu \beta \beta$ half-life and LHC bound on $\Delta^{\pm \pm}_R$ mass. The mass of $W_R$ boson is varied in the range $M_{W_R} \in 3-100$ TeV.}
\label{Thalf2}
\end{figure}
\begin{figure}[htb]
\centering
\includegraphics[scale=0.7]{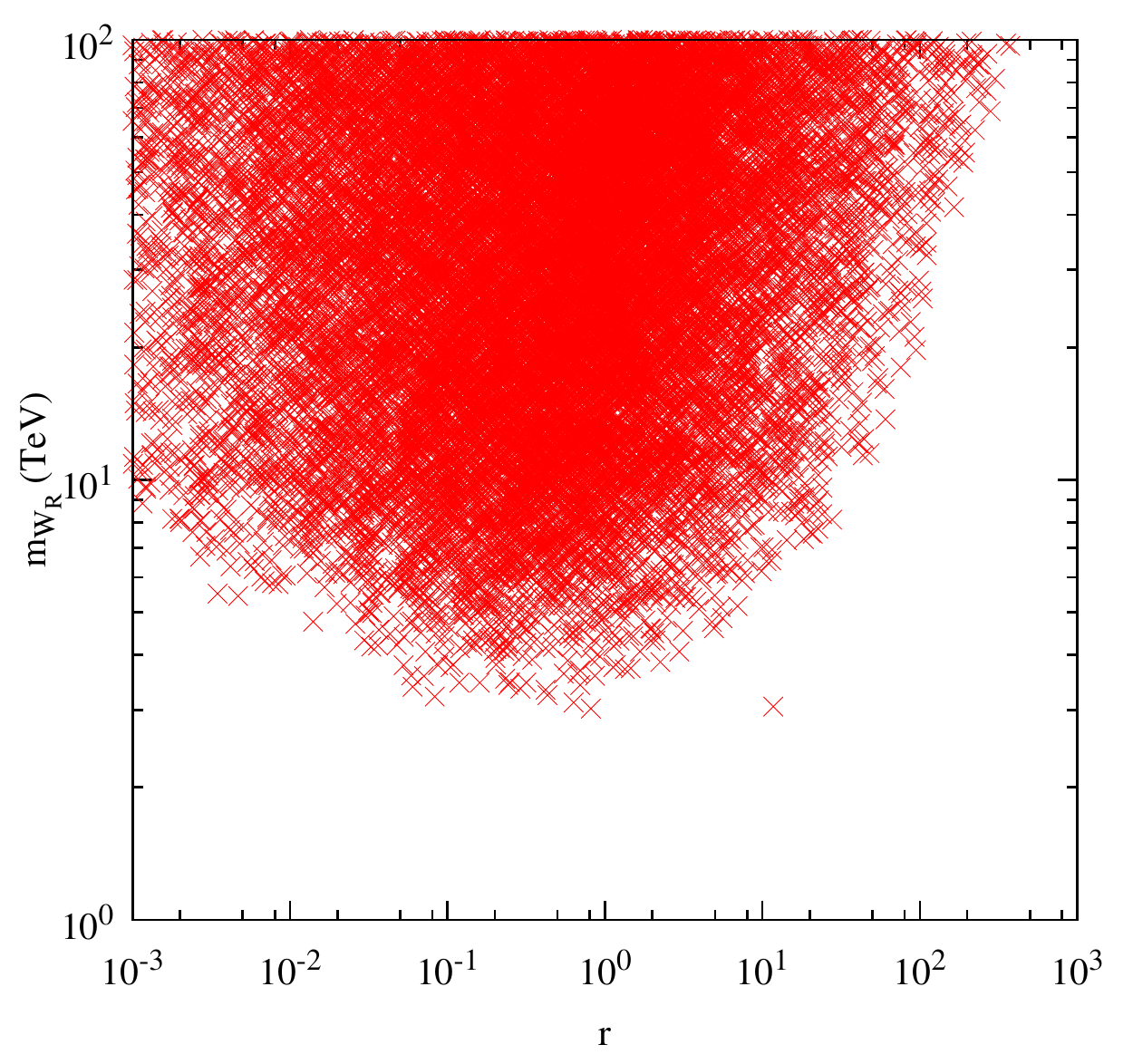}
\caption{Allowed Parameter space in heavy gauge boson mass $M_{W_R}$ versus $r=\frac{M_N}{M_{\Delta}}$, the ratio of the masses of heavy neutrino and that of the doubly charged scalar $\Delta^{\pm \pm}_R$ from KamLAND-Zen bound on $0\nu \beta \beta$ half-life and LHC bound on $\Delta^{\pm \pm}_R$ mass. The relevant masses are varied in the ranges: $M_{W_R} \in 3-100$ TeV, $M_N \in 1-2\times 10^5$ GeV, $M_{\Delta^{\pm \pm}_R} \in 420-2\times 10^5$ GeV.}
\label{Thalf3}
\end{figure}

\section{Neutrinoless Double Beta Decay}
\label{sec3}
Although the active neutrino masses are dominantly of Dirac type, the model discussed above can still give rise to lepton number violating processes due to the presence of additional gauge bosons and heavy Majorana fermions. The leading contributions to $0\nu \beta \beta$ process is shown in terms of the Feynman diagrams in figure \ref{ndbd1}. The $W_L$ mediated diagrams will be suppressed by the tiny Majorana masses of the left handed neutrinos. The mixed $W_L-W_R$ diagrams are also suppressed due to the tiny mixing between $\nu_L$ and $N_R$. The first diagram in figure \ref{ndbd1} correspond to the triplet scalar $\Delta_R$ mediated process whose contribution to the $0\nu \beta \beta$ amplitude is given by 
\begin{equation}
A_{R\Delta} \propto G^2_F \left ( \frac{M_{W_L}}{M_{W_R}} \right )^4 \sum_i \frac{V^2_{ei} M_i}{M^2_{\Delta^{--}_R}}
\end{equation}
where $V$ is approximately equal to the diagonalising matrix of the heavy neutrino mass matrix $M_{RR}$ and $M_i$ are the mass eigenvalues of $M_{RR}$. The left-handed counterpart of this process where $W_R, \Delta_R$ are replaced by $W_L, \Delta_L$ does not exist in this particular model. The contribution from the heavy neutrino and $W_R$ exchange (second Feynman diagram in figure \ref{ndbd1}) can be written as 
\begin{equation}
A_{NRR} \propto G^2_F \left ( \frac{M_{W_L}}{M_{W_R}} \right )^4 \sum_i \frac{V^{*2}_{ei}}{M_i} 
\end{equation}
Combining these two dominant contributions, the half-life of $0\nu \beta \beta$ process can be written as 
\begin{equation}
\frac{1}{T^{0\nu}_{1/2}} = G^{0\nu}_{01} \bigg ( \lvert \mathcal{M}^{0\nu}_N (\eta^R_N+\eta_{\Delta_R}) \rvert^2 \bigg )
\label{eq:halflife}
\end{equation}
where 
$$ \eta^R_N= m_p\left ( \frac{M_{W_L}}{M_{W_R}} \right )^4 \sum_i \frac{V^{*2}_{ei}}{M_i}, \;\;\;\; \eta_{\Delta_R}= m_p\left ( \frac{M_{W_L}}{M_{W_R}} \right )^4 \sum_i \frac{V^2_{ei} M_i}{M^2_{\Delta^{--}_R}} $$
Here $m_p$ is the proton mass and $\mathcal{M}$ are nuclear matrix elements (NME) whereas $G^{0\nu}_{01}$ is the phase space factor. The numerical values of NME and the phase space factor are shown in table \ref{tableNME} for different nuclei.  Here, we consider a general structure of $V$, vary the masses heavy neutrinos from 1 GeV to $v_R \sim v_{\delta_R} \sim 6$ TeV while keeping $\Delta^{\pm \pm}_R$ mass in the 420 GeV to 6 TeV range, and plot $T^{0\nu}_{1/2}$ as a function of $r = m_N/m_{\Delta}$, the ratio between the heaviest among the heavy neutrinos and the doubly charged scalar mass. For equal left-right gauge couplings $g_L = g_R$, this corresponds to $M_{W_R} \approx 3$ TeV. The variation of half-life is shown in figure \ref{Thalf}. The resulting half-life is then compared against the latest experimental bounds. For example, the recent bound from the KamLAND-Zen experiment constrains $0\nu \beta \beta$ half-life \cite{kamland2}
$$ T^{0\nu}_{1/2} (\text{Xe}136) > 1.1 \times 10^{26} \; \text{yr} $$
Similarly, the GERDA experiment has also reported a slight improvement over their earlier estimates and reported the half-life to be \cite{GERDA2}
\begin{equation}
T^{0\nu}_{1/2} (\text{Ge}76) > 4.0 \times 10^{25} \; \text{yr}
\end{equation}
It can be seen from the plot in figure \ref{Thalf} that the latest experimental bounds still allow $r \sim 1-2$. The sharp cut near $r \sim 1-2$ results from including the LHC lower bound on $\Delta^{\pm \pm}_R$ mass ($~420$ GeV). To see the allowed parameter space more clearly, we also show the doubly charged scalar mass $m_{\Delta^{\pm \pm}_R}$ versus heavy neutrino mass $m_{N_R}$ allowed from $0\nu\beta\beta$ and LHC limits in figure \ref{Thalf2}. Similar allowed parameter space is shown for $M_{W_R}$ against $r=\frac{M_N}{M_{\Delta}}$ in figure \ref{Thalf3}.
\begin{center}
\begin{table}[htb]
\begin{tabular}{|c|c|c|}
\hline
Isotope & $G^{0\nu}_{01} \; (\text{yr}^{-1})$  & $\mathcal{M}^{0\nu}_N$  \\
\hline
$ \text{Ge}-76$ & $5.77\times10^{-15}$  & $233-412$  \\
$ \text{Xe}-136$ & $3.56 \times 10^{-14}$  & $164-172$  \\
\hline
\end{tabular}
\caption{Values of phase space factor and nuclear matrix elements used in the analysis.}
\label{tableNME}
\end{table}
\end{center}
\begin{figure}[htb]
\centering
\includegraphics[scale=0.75]{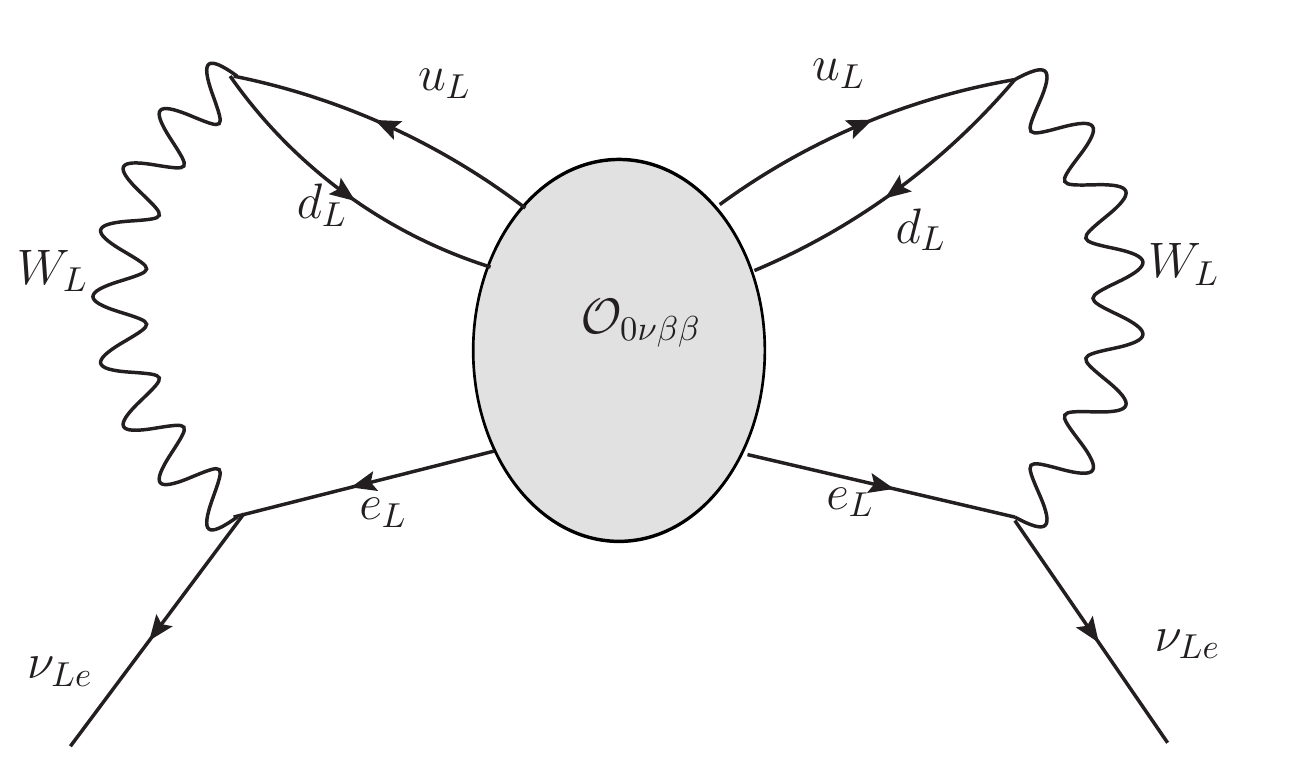}
\caption{Four loop contribution to the Majorana mass of light neutrinos through the Butterfly diagram.}
\label{ndbd2}
\end{figure}
\begin{figure}[htb]
\centering
\includegraphics[scale=1.0]{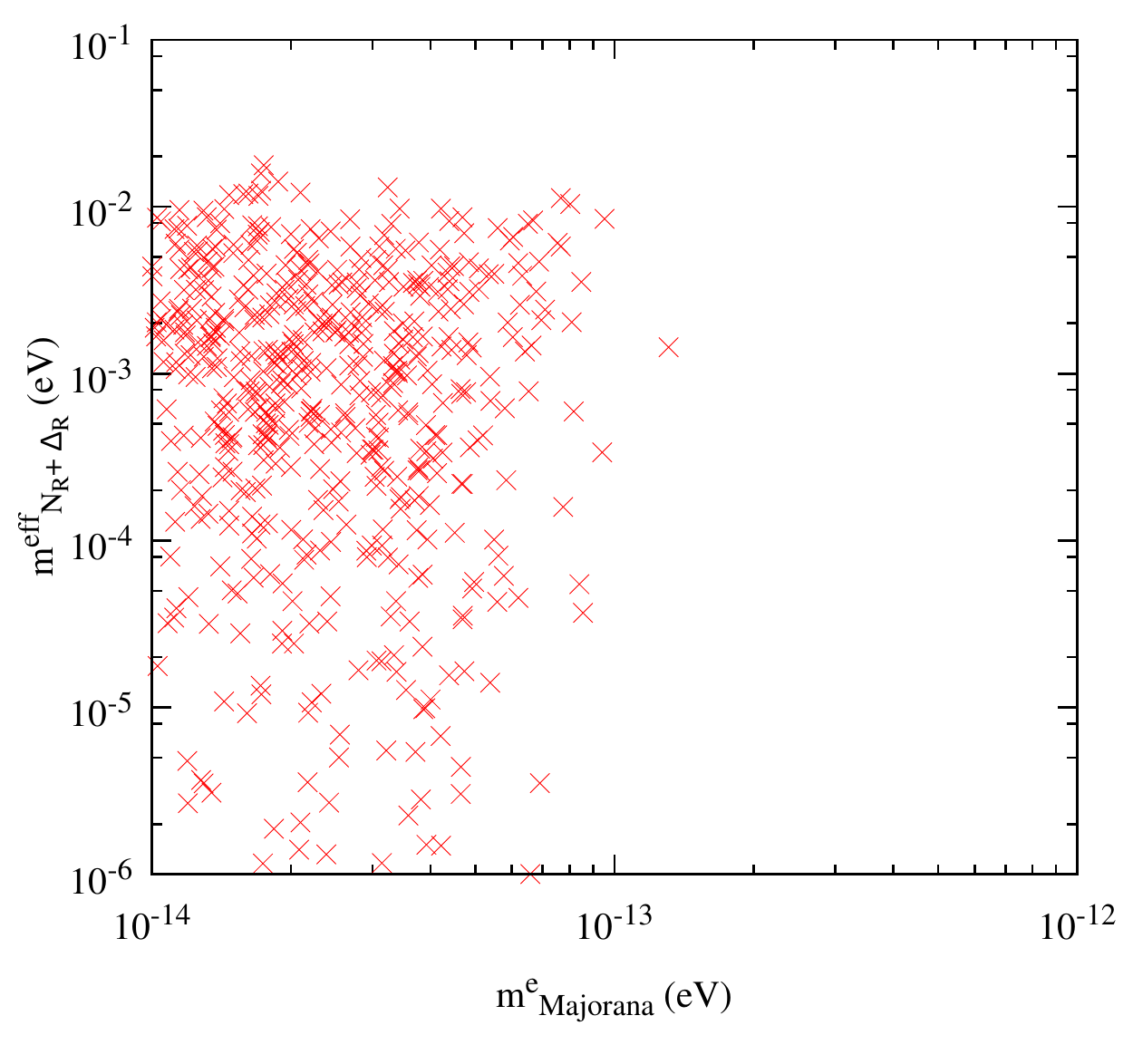}
\caption{Effective Majorana mass responsible for $0\nu\beta \beta$ versus Majorana mass of electron type neutrino originating from type I seesaw. The values of $m^{\text{eff}}$ satisfy the latest experimental bounds on $0\nu\beta \beta$ half-life. The relevant masses are varied in the ranges: $M_{W_R} \in 3-100$ TeV, $M_N \in 1-2\times 10^5$ GeV, $M_{\Delta^{\pm \pm}_R} \in 420-2\times 10^5$ GeV.}
\label{meevsMajorana}
\end{figure}

As mentioned earlier, the Schechter-Valle theorem \cite{schvalle} implies that any non-zero amplitude of $0\nu\beta \beta$ induces a non-zero effective Majorana mass to the electron type neutrino, irrespective of the underlying mechanism behind the $0\nu\beta \beta$ process. The lowest possible order such a mass term can arise is through the four loop diagram shown in figure \ref{ndbd2} which was computed by \cite{4loopcomp}. The blob in the Feynman diagram shown in figure \ref{ndbd2} indicates the absence of any a priori knowledge about the underlying mechanism responsible for $0\nu\beta \beta$. Depending on the underlying mechanism, the helicities of the quarks and electrons will also be different. However, to complete the four loop diagram with two left handed neutrinos in the external fermion legs, one must incorporate the standard left-handed gauge interactions, as shown in figure \ref{ndbd2}. In case the charged fermions taking part in $0\nu\beta \beta$ are of opposite helicities (like in the present model, where the quarks and electrons taking part in $0\nu \beta \beta$ are right handed), necessary mass insertions should be made to make them couple to $W_L$ bosons. The authors in \cite{4loopcomp} showed all possible Lorentz invariant operators that can contribute to $0\nu\beta \beta$ and showed that one such operator contributes a maximum of 
$$ \delta M^{ee}_{\nu} \approx (0.74-5) \times 10^{-28} \; \text{eV}$$
to the Majorana mass of electro type neutrino. It was referred to as "maximum" contribution because the upper limit on $0\nu \beta \beta$ amplitude from latest experiments was incorporated. Thus, it does not conflict with the validity of the Schechter-Valle theorem which guarantees a minimum non-zero contribution to the Majorana mass of electron type neutrino, if there is a non-zero $0\nu \beta \beta$ amplitude. This confirms the qualitative validity of the Schechter-Valle theorem, though the calculated Majorana mass term is way too small compared to the neutrino mass squared differences. Although in our model, we know the helicities of the charged fermions taking part in $0\nu \beta \beta$, we do not calculate the Majorana mass term induced by this decay at four or higher loop orders, as we already have a more dominant contribution to neutrino Majorana mass terms through type I seesaw discussed above. Since all Majorana type contribution to light neutrino masses are highly suppressed in this model, the light neutrinos remain predominantly Dirac in spite of observable lepton number violation through $0\nu \beta \beta$. Quantitatively, we show the difference between effective Majorana mass appearing in $0\nu \beta \beta$ and Type I seesaw contribution to the Majorana mass of electron type neutrino in the plot shown in figure \ref{meevsMajorana}. The effective Majorana mass corresponding to the two major contributions to the $0\nu \beta \beta$ is 
$$ m^{\text{eff}}_{N_R+\Delta_R} = \lvert  m^{\text{eff}}_{N_R} + m^{\text{eff}}_{\Delta_R} \big \rvert$$ 
where 
$$ m^{\text{eff}}_{N_R} = p^2 \frac{M^4_{W_L}}{M^4_{W_R}} \frac{V^{*2}_{ei}}{M_i}, \quad m^{\text{eff}}_{\Delta_R}=p^2 \frac{1}{M^4_{W_R}} \frac{V^2_{Rei}M_i}{M^2_{\Delta_R}}$$
with $p \sim 100$ MeV being the typical momentum exchange of the process. It is clear from the figure \ref{meevsMajorana} that the effective Majorana mass for $0\nu \beta \beta$ can be within the current experimental sensitivity while the Majorana mass of light neutrinos remain many order of magnitudes smaller than observed neutrino masses.
\begin{figure}[htb]
\centering
\includegraphics[scale=0.50]{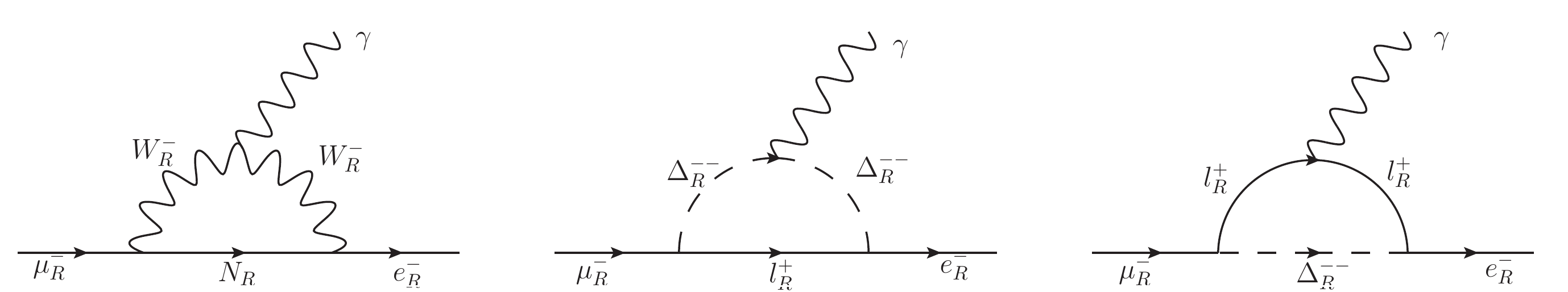}
\caption{New physics contribution to LFV decays}
\label{lfv1}
\end{figure}
\begin{figure}[htb]
\centering
$
\begin{array}{cc}
\includegraphics[scale=0.7]{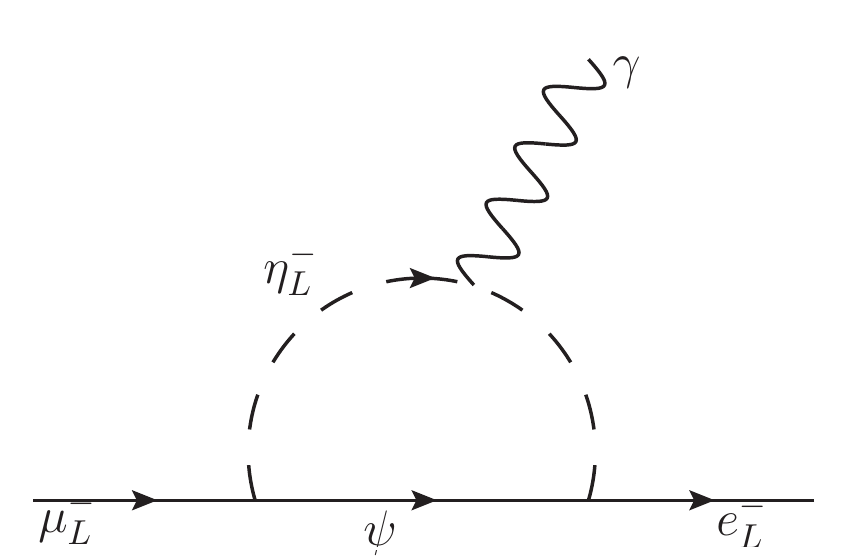} &
\includegraphics[scale=0.8]{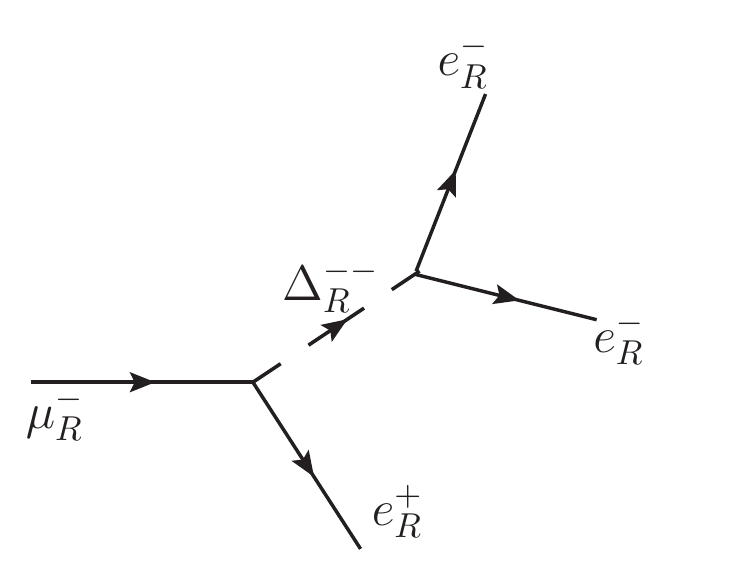} \\
\end{array}$
\caption{New physics contribution to LFV decays}
\label{lfv2}
\end{figure}
\begin{figure}[htb]

\centering
$
\begin{array}{cc}
\includegraphics[scale=0.6]{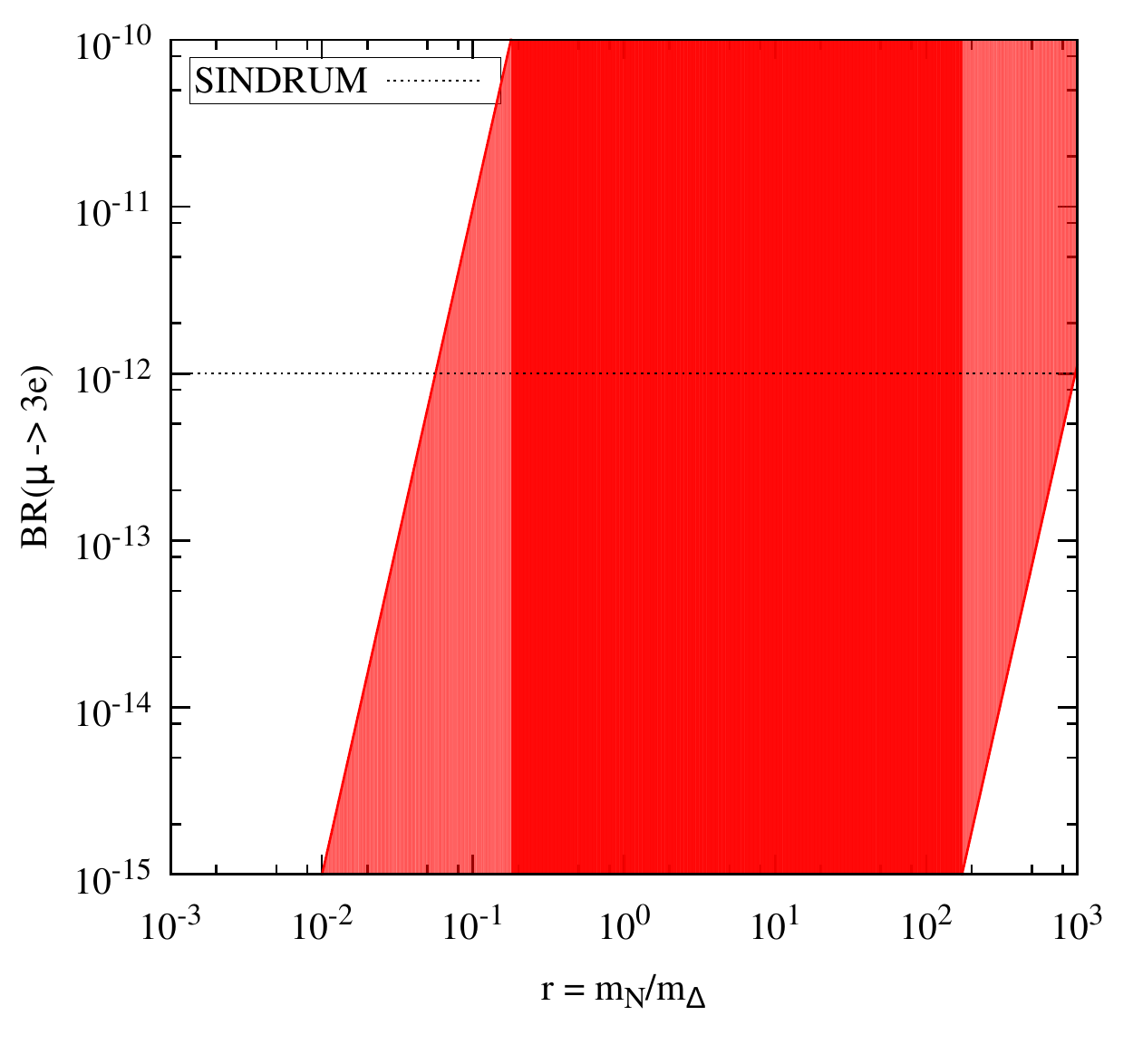} &
\includegraphics[scale=0.6]{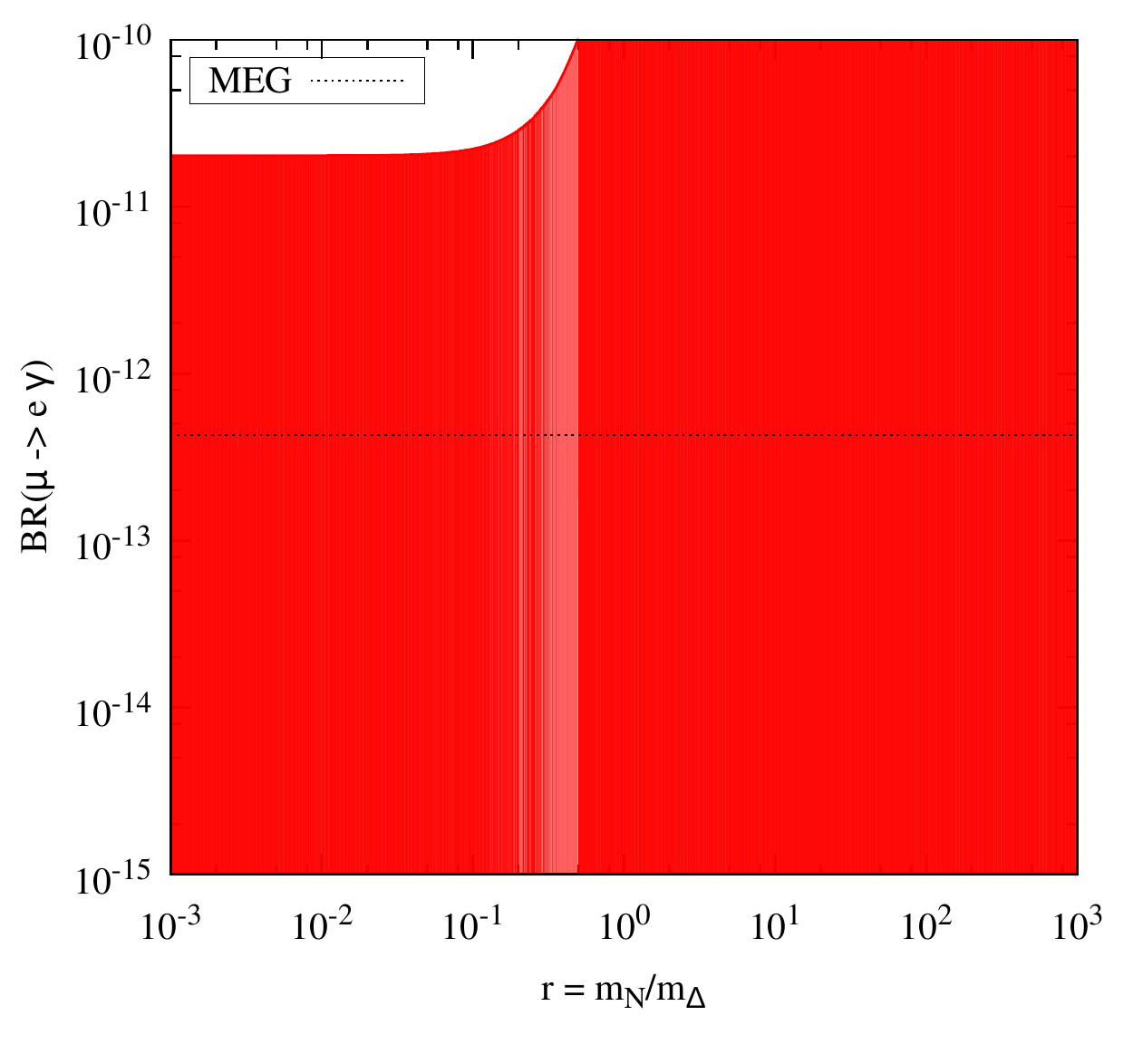} \\
\end{array}$
\caption{Branching ratio for $\mu \rightarrow 3e, \mu \rightarrow e \gamma$ as a function of $r=\frac{M_N}{M_{\Delta}}$, the ratio of the masses of heavy neutrino and that of the doubly charged scalar from the triplet $\Delta_R$. The relevant masses are varied in the ranges: $M_{W_R} \in 3-100$ TeV, $M_N \in 1-2\times 10^5$ GeV, $M_{\Delta^{\pm \pm}_R} \in 420-2\times 10^5$ GeV.}
\label{clfv1}
\end{figure}

\begin{figure}[htb]
\centering
\includegraphics[scale=1.0]{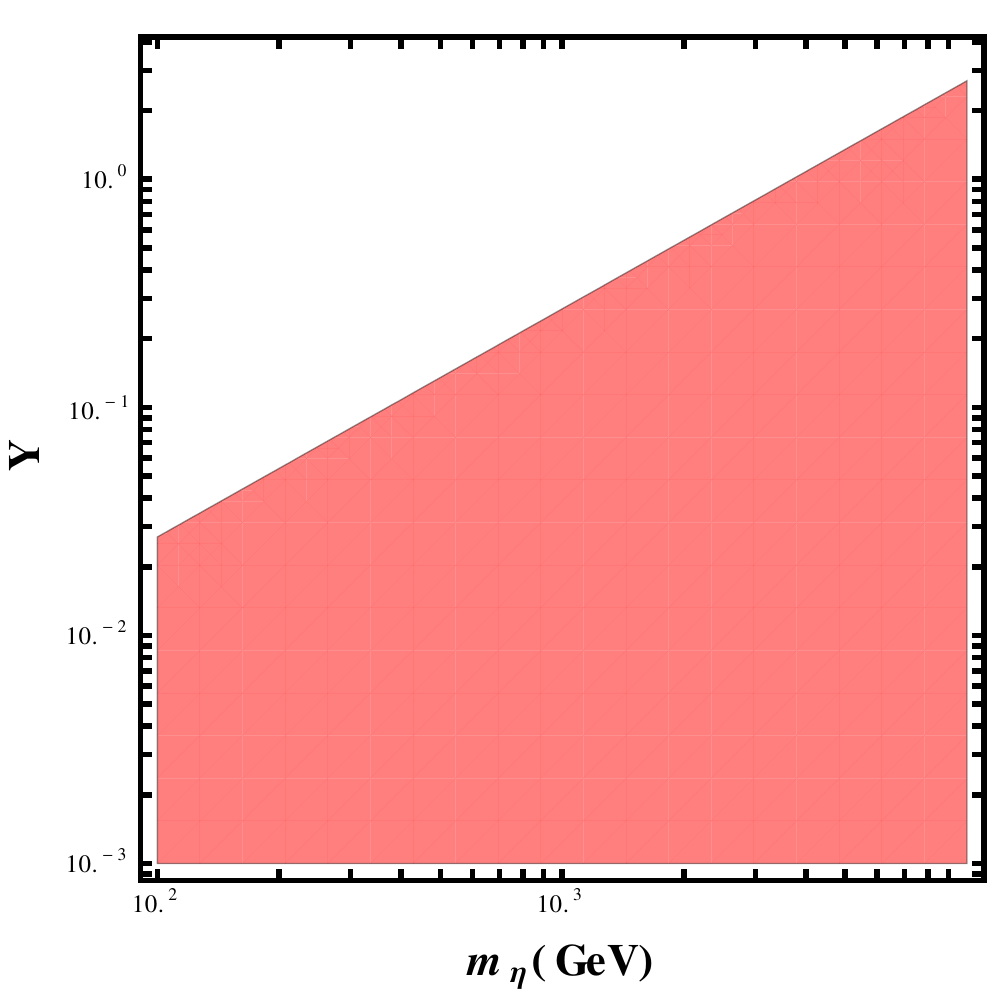}
\caption{Contribution from the charged component of the left handed scalar doublet $\eta_L$ to the $\mu \rightarrow e \gamma$ decay assuming $Y^{\mu}_{\nu} = Y^e_{\nu} = Y, m_{\psi} = 10m_{\eta_L}$.}
\label{clfv2}
\end{figure}

\section{Charged Lepton Flavour Violation}
\label{sec4}
Charged lepton flavour violating processes which remain suppressed in the SM, could get significantly enhanced in the presence of BSM physics around the TeV corner and can be probed at ongoing or near future experiments. Here we consider the new physics contributions to $\mu \rightarrow e\gamma$ as well as $\mu \rightarrow 3e$ mediated by charged scalars, right handed vector boson $W_R$ and heavy fermions $N_R$ as seen from the Feynman diagrams shown in figure \ref{lfv1} and \ref{lfv2}. The latest bound from the MEG collaboration is $\text{BR}(\mu \rightarrow e \gamma) < 4.2 \times 10^{-13}$ at $90\%$ confidence level \cite{MEG16}. Similarly, the SINDRUM collaboration has put bound on the other LFV decay process $\text{BR}(\mu \rightarrow 3e) < 1.0 \times 10^{-12}$ \cite{sindrum}. The contribution from the diagrams in figure \ref{lfv1} to $\mu \rightarrow e\gamma$ is given by \cite{ndbd10}
\begin{equation}
\text{BR}(\mu \rightarrow e\gamma) =\frac{3\alpha_{\text{em}}}{2\pi} \left (  \lvert G^{\gamma}_R \rvert^2 \right )
\label{eqBR1}
\end{equation}
where $\alpha_{\text{em}}=e^2/4\pi$ and the form factors $G^{\gamma}_{R}$ are given by
\begin{equation}
G^{\gamma}_R = \sum^3_{i=1}   \bigg ( (V)_{\mu i}(V)^*_{ei} \bigg[\frac{M^2_{W_L}}{M^2_{W_R}}G^{\gamma}_1 (y_i)+\frac{2y_i}{3} \frac{M^2_{W_L}}{M^2_{\Delta^{++}_R}} \bigg]  \bigg)
\end{equation}
In the above expressions $y_i\equiv (M_i/M_{W_R})^2$. The loop functions $G^{\gamma}_{1}$ are given by
\begin{equation}
G^{\gamma}_1 (a) = -\frac{2a^3+5a^2-a}{4(1-a)^3}-\frac{3a^3}{2(1-a)^4} \ln{a}
\nonumber
\end{equation}
On the other hand, the first diagram in figure \ref{lfv2} contributes to the decay width of $\mu \rightarrow e \gamma$ as 
\begin{align}
\Gamma (\mu \rightarrow e \gamma) &= \frac{Y^2_{\nu} Y^2_r \left(m^2_\mu - m^2_e \right)^3(m^2_\mu + m^2_e)}{4096 \pi^5 m^3_\mu m^4_{\eta^-_L}} \left[\frac{ \left((t-1)(t(2t+5)-1) + 6t^2\ln t\right)^2}{144 (t-1)^8}\right]
\end{align}
where $t = m^2_{\psi_i}/m^2_{\eta^-_L}$. The corresponding branching ratio can be found by
$$ \text{BR}(\mu \rightarrow e \gamma) \approx \frac{\Gamma (\mu \rightarrow e \gamma)}{\Gamma_{\mu}} $$
where $\Gamma_{\mu} \approx 2.996 \times 10^{-19}$ GeV denotes the total decay width of muon.

The second diagram shown in figure \ref{lfv2} contributes to the LFV process $\mu \rightarrow 3e$ mediated by doubly charged boson $\Delta^{++}_R$ as \cite{LFVLR}
\begin{equation}
\text{BR}(\mu \rightarrow 3e) = \frac{1}{2} \lvert h_{\mu e} h^{*}_{ee} \rvert^2 \left (\frac{M^4_{W_L}}{M^4_{\Delta^{++}_R}} \right )
\label{eqBR}
\end{equation}
where the couplings $h$ are given by
\begin{equation}
h_{ij} = \sum_n \left ( V \right )_{ni} \left ( V \right )_{nj} \sqrt{\left(\frac{M_i}{M_{W_R}}\right)^2}
\label{eqhij}
\end{equation}
Since the heavy neutrino mass matrix $M_{RR}$ is not related to the leading order light neutrino mass, we can parametrise it independently as $M_{RR}=V M^{(\text{diag})}_{RR} V^T$. Here $M^{(\text{diag})}_{RR}=\text{diag}(M_1, M_2, M_3)$ is the diagonal light neutrino mass matrix. The $3\times 3$ mixing matrix $V$ can be parametrised in a way similar to the Pontecorvo-Maki-Nakagawa-Sakata (PMNS) leptonic mixing matrix in terms of three mixing angles ($\phi_{ij}; i, j=1,2,3$) and three phases ($ \delta, \alpha, \beta $). We show the new physics contribution to these LFV decays as a function of $r = m_N/m_{\Delta}$ in figure \ref{clfv1}. It can be seen that the latest experimental bounds still allows large values of $r$ beyond the ones allowed by the constraints from $0\nu \beta \beta$ experiments. We also calculate the contribution from $\eta_L$ mediated diagram in figure \ref{lfv2} to $\mu \rightarrow e \gamma$ by assuming $Y^{\mu}_{\nu} = Y^e_{\nu} = Y, m_{\psi} = 10m_{\eta_L}$. The region of parameter space satisfying the latest MEG bound \cite{MEG16} is shown in figure \ref{clfv2}. We choose a heavier $\psi$ than $\eta_L$ as we intend to discuss scalar dark matter in the next section. Moreover, a heavy Dirac fermion $\psi$ mediating such loop diagrams can also give rise to Dirac leptogenesis as discussed recently by \cite{dbad1}.

\section{Dark Matter}
\label{sec5}
Several astrophysical and cosmological evidences suggest the presence of dark matter (DM) in our Universe. The latest data collected by the Planck experiment suggests around $26\%$ of the present Universe's energy density being made up of dark matter \cite{Planck15}. Their estimate can also be expressed in terms of density parameter $\Omega$ as
\begin{equation}
\Omega_{\text{DM}} h^2 = 0.1187 \pm 0.0017
\label{dm_relic}
\end{equation}
where $h = \text{(Hubble Parameter)}/100$ is a parameter of order unity. According to the list of criteria, a dark matter candidate must fulfil \cite{bertone}, none of the SM particles can qualify for it. Interestingly, the model we are studying in this work, provides several dark matter candidates. The dark matter in the model is in fact, a combination of scotogenic dark matter \cite{m06} and minimal left-right dark matter (MLRDM) formalism \cite{Heeck:2015qra,Garcia-Cely:2015quu}. In the scotogenic scenario, the lightest particle in the internal lines of the one loop diagram for neutrino mass is a stable dark matter candidate. In our model, the list of such particles include $\eta^0_L, \psi, \chi_1$. Here we consider the $\eta^0_L$ as DM due to the better detection prospects by virtue of its gauge interactions. On the other hand, in the MLRDM formalism, stable dark matter candidates arise accidentally due to the appropriate choices of their $SU(2)$ dimensions, in the spirit of minimal dark matter framework \cite{Cirelli:2005uq,Garcia-Cely:2015dda,Cirelli:2015bda}. This includes $\eta^0_L, \eta^0_R$ in our model. This scenario was in fact studied in \cite{Garcia-Cely:2015quu} where a pair of scalar doublets $\eta_{L,R}$ were added to the minimal LRSM. However, in minimal LRSM, there exists a coupling $\eta^T_L \Phi \eta_R$ with $\Phi$ being the scalar bidoublet. This leads to the decay of the heavier DM into the lighter one and SM fermions mediated by the Higgs. In the present model, the chosen discrete symmetries do not allow any renormalisable coupling between $\eta_L$ and $\eta_R$ leading to the tantalising possibility of multi-component DM where both of them can contribute to the total dark matter relic abundance. Unlike in \cite{Heeck:2015qra,Garcia-Cely:2015quu}, it is not stabilised by the $Z_2 = (-1)^{B-L}$ subgroup of the $U(1)_{B-L}$ gauge group as it is broken already by the vev of the neutral components of the scalar doublets $H_{L,R}$ which are odd under this $Z_2$ symmetry. The dark matter candidates in our model are stable accidentally due to absence of renormlisable operator leading to their decay, similar to the minimal dark matter formalism. If we consider higher dimensional operators, it is possible to generate decay diagrams responsible for dark matter decay. For example, dimension five operators like $(\eta_L \eta_R H^{\dagger}_L H^{\dagger}_R \chi_3)/\Lambda$ can lead to heavier dark matter (say $\eta^0_R$) decay into the lighter one ($\eta^0_L$). Similarly, the lighter dark matter can also decay through higher dimensional operators like $(\eta^{\dagger}_L H_L \chi_2)^2 \chi_3/\Lambda^3, (\eta_L H_L \Delta_L \chi_2)^2 \chi_3/\Lambda^5$ and so on. Constraints on dark matter lifetime will put lower limits on this cut-off scale $\Lambda$, details of which can be found elsewhere.

The relic abundance calculation of scalar doublet DM $\eta^0_{L, R}$ is similar to that of inert doublet model (IDM) studied extensively in the literature \cite{dbad1,m06,Barbieri:2006dq,Majumdar:2006nt,LopezHonorez:2006gr,ictp,borahcline, honorez1,DBAD14}. However, their individual contributions to total DM abundance is different due to their different gauge interactions. The authors of \cite{Garcia-Cely:2015quu} considered only the gauge interactions of $\eta^0_L$ and $\eta^0_R$ such that both of them can be stable and their relic abundances can be calculated independently, in the absence of zero left-right mixing. They showed that for $M_{W_R} = 2$ TeV, only $m_{\eta^0_L} = m_{\eta^0_R} \approx 150$ GeV satisfies the total DM relic abundance constraint. However, if we turn on other interactions, then more allowed parameter space should come out. In this work, we consider the interactions of $\eta^0_L$ with the Higgs boson whereas restrict the dominant interactions of $\eta^0_R$ to the gauge sector only. The present model allows both $\eta^0_L, \eta^0_R$ to be stable even if we turn on all possible interactions, which was not the case in minimal LRSM discussed by \cite{Garcia-Cely:2015quu}. For simplicity, we keep the $\eta^0_R$-Higgs interaction is almost switched off in order to keep the relic abundance calculations of two DM candidates independent of each other. This will become clear from the following discussion.

The relic abundance of a DM particle is calculated by solving the Boltzmann equation
\begin{equation}
\frac{dn_{\chi}}{dt}+3Hn_{\chi} = -\langle \sigma v \rangle (n^2_{\chi} -(n^{\text{eqb}}_{\chi})^2)
\label{BE1}
\end{equation}
where $n_{\chi}$ is the dark matter number density and $n^{eqb}_{\chi}$ is the corresponding equilibrium number density. $H$ is the Hubble expansion rate of the Universe and $ \langle \sigma v \rangle $ is the thermally averaged annihilation cross section of the dark matter particle $\chi$. In terms of partial wave expansion $ \langle \sigma v \rangle = a +b v^2$. Clearly, in the case of thermal equilibrium $n_{\chi}=n^{\text{eqb}}_{\chi}$, the number density is decreasing only by the expansion rate $H$ of the Universe. The approximate analytical solution of the above Boltzmann equation gives \cite{Kolb:1990vq, kolbnturner}
\begin{equation}
\Omega_{\chi} h^2 \approx \frac{1.04 \times 10^9 x_F}{M_{Pl} \sqrt{g_*} (a+3b/x_F)}
\end{equation}
where $x_F = m_{\chi}/T_F$, $T_F$ is the freeze-out temperature, $g_*$ is the number of relativistic degrees of freedom at the time of freeze-out and $M_{Pl} \approx 10^{19}$ GeV is the Planck mass. Here, $x_F$ can be calculated from the iterative relation 
\begin{equation}
x_F = \ln \frac{0.038gM_{\text{Pl}}m_{\chi}<\sigma v>}{g_*^{1/2}x_F^{1/2}}
\label{xf}
\end{equation}
The thermal averaged annihilation cross section $\langle \sigma v \rangle$ is given by \cite{Gondolo:1990dk}
\begin{equation}
\langle \sigma v \rangle = \frac{1}{8m^4_{\chi}T K^2_2(m_{\chi}/T)} \int^{\infty}_{4m^2_{\chi}}\sigma (s-4m^2_{\chi})\surd{s}K_1(\surd{s}/T) ds
\label{eq:sigmav}
\end{equation}
where $K_i$'s are modified Bessel functions of order $i$, $m_{\chi}$ is the mass of Dark Matter particle and $T$ is the temperature. In the presence of multiple DM candidates, we have multiple Boltzmann equations similar to the one in \eqref{BE1}. Usually, these multiple Boltzmann equations are coupled due to the fact that one DM candidate can self-annihilate into another and vice versa. However, if we turn off the interactions mediating different DM candidates, then these equations become decoupled and hence can be solved independently. We keep them decoupled in our work simply by assuming negligible $\eta^0_R$-Higgs couplings and quartic couplings between $\eta_L, \eta_R$. These couplings can not be forbidden by the underlying discrete symmetries. Since the left-right mixing is also negligible (vanishing at tree level), there exists no annihilation channels of $\eta^0_R$ type DM to $\eta^0_L$ and vice versa. The couplings between $\eta_{L,R}$ and the Higgs also help in splitting the masses between charged and neutral components of the scalar doublets. This can occur through scalar interactions like this
\begin{equation}
\mathcal{L} \supset  \lambda_{5L,R} (H^{\dagger i}_{L,R}H_{L,R i})(\eta^{\dagger j}_{L,R}\eta_{L,R j})+\lambda_{6 L,R} 
(H^{\dagger i}_{L,R} H_{L,R j})(\eta^{\dagger j}_{L,R}\eta_{L,R i})
\label{scalarpot2}
\end{equation}
This along with the parts of scalar Lagrangian given in equation \eqref{scalarL} gives us the physical masses of $\eta_{L,R}$ components at tree level. They are given by
\begin{eqnarray}
m^2_{\eta_{Ls}} &=& \mu^2_{\eta_L}+ \frac{1}{2}(\lambda_{5L} v^2_L + \lambda_{6L} v^2_L + \lambda_4 v_{\delta_L} u_3) \nonumber\\
m^2_{\eta_{Lp}} &=& \mu^2_{\eta_L}+ \frac{1}{2}(\lambda_{5L} v^2_L + \lambda_{6L} v^2_L - \lambda_4 v_{\delta_L} u_3) \nonumber\\
m^2_{\eta^{\pm}_L} &=& \mu^2_{\eta_L}+ \frac{1}{2}\lambda_{5L} v^2_L  \nonumber\\
m^2_{\eta_{Rs}} &=& m^2_{\eta_{Rp}} = \mu^2_{\eta_R}+ \frac{1}{2}(\lambda_{5R} v^2_R + \lambda_{6R} v^2_R ) \nonumber\\
m^2_{\eta^{\pm}_R} &=& \mu^2_{\eta_R}+ \frac{1}{2}\lambda_{5R} v^2_R
\label{treemass}
\end{eqnarray}
where we are ignoring the possible quartic couplings between L and R sectors. It can be seen that the neutral scalar and pseudoscalar of $\eta_L$ acquire a tree level mass split due to the vev of $\Delta_L, \chi_3$. Similarly there is a mass splitting between charged and neutral component making sure that the neutral component can be lighter and hence a dark matter candidate. The scalar and pseudoscalar components of $\eta_R$ however remains degenerate at tree level.

The relic abundance of $\eta^0_L$ is calculated in a way similar to the IDM. Since this is a complex field, one can write it as $\eta^0_L = (\eta^0_{Ls} + i \eta^0_{Lp})/\sqrt{2} $. From the scalar Lagrangian \eqref{scalarL}, \eqref{scalarpot2}, it can be seen that the real and imaginary components of $\eta^0_L$ have a mass degeneracy in the absence of the triplet scalar $\Delta_L$. Due to the quartic term $\lambda_4 \eta_L \eta_L \Delta_L \chi_3$, non-zero vev's of the neutral component of $\Delta_L$ and $\chi_3$ break the mass degeneracy of $\eta^0_{Ls, Lp}$. This is necessary to evade large inelastic DM-nucleon scattering at direct detection experiments due to $\eta^0_{Ls, Lp}-Z_L$ couplings. Taking the typical kinetic energy of a dark matter particle to be approximately 100 keV, one can obtain the constraint on the mass splitting as
\begin{equation}
|m^2_{\eta_{Lp}} -m^2_{\eta_{Ls}}|= 2 \lambda_4 \langle \chi_3 \rangle v_{\delta_L} >  (m_{\eta_{Lp}} +m_{\eta_{Ls}}) \times 100 \; \text{keV} 
\end{equation}
Considering the maximum possible value of $v_{\delta_L} (\sim 2 \; \text{GeV})$ allowed by the constraints on the $\rho$ parameter discussed earlier, we get the following constraint
\begin{equation}
\lambda_4 \langle \chi_3 \rangle > (m_{\eta_{Lp}} +m_{\eta_{Ls}}) \times 2.5 \times 10^{-5}  \; \text{GeV} 
\end{equation}
which can be achieved naturally for the region of parameter space discussed in this work. A large mass splitting also makes the effects of coannihilation between different components of the $\eta_L$ doublet negligible. On the other hand, there is no such term in the Lagrangian that can lift the mass degeneracy between scalar and pseudoscalar parts of $\eta^0_R$ DM. This is however, not as problematic as having a degeneracy in the $\eta^0_L$ case, as the corresponding neutral boson $Z_R$ is much heavier to suppress the inelastic DM-nucleon scattering. In the absence of non-gauge interactions of $\eta_R$, the mass splitting between the charged and neutral components of $\eta_R$ also remain zero, at least at tree level. At one loop level however, there arises a mass splitting between $\eta^{\pm}_R$ and $\eta^0_R$ given in \cite{Garcia-Cely:2015quu} as
\begin{equation}
M_Q-M_0 = \frac{M}{16 \pi^2} \left ( \sum_V g^2_{V, 0} g(r_V) - \sum_V g^2_{V, Q} g(r_V) \right )
\end{equation}
$g_{V, X}$ is the vector boson coupling to the scalar and the loop function $g(r)$ is given by
$$ g(r) =-5-\frac{r}{4} \left ( 2r^3 \log{r} + (r^2-4)^{3/2} \log{\frac{r^2-2-r\sqrt{r^2-4}}{2}} \right )$$
with $r_V = M_V/M$. Here $M_V$ is the mass of the vector boson and $M$ is the tree level degenerate mass of the $\eta_R$ components. To avoid the issue of divergence of renormalisibility involved in such loop corrections, here we simply assume a tree level mass splitting of 1 GeV between $m_{\eta_{R}}$ and $m_{\eta^{\pm}_{R}}$. From the tree level masses given in equation \eqref{treemass}, it can be seen that such a mass splitting can arise by appropriately choosing the quartic coupling $\lambda_{6R}$. Since $v_R$ is large, of TeV order, even a tiny $\lambda_{6R}$ can generate such a splitting, without introducing any new dominant annihilation channels of $\eta_R$ dark matter. For such mass splittings, coannihilation effects may be important while calculating the relic abundance of $\eta^0_R$ DM. Such effects were studied by several groups in \cite{Griest:1990kh, coann_others}. Here we incorporate the effects of coannihilation in relic abundance calculations, following the framework given by \cite{Griest:1990kh}.
\begin{figure}[htb]
\centering
\includegraphics[scale=1.0]{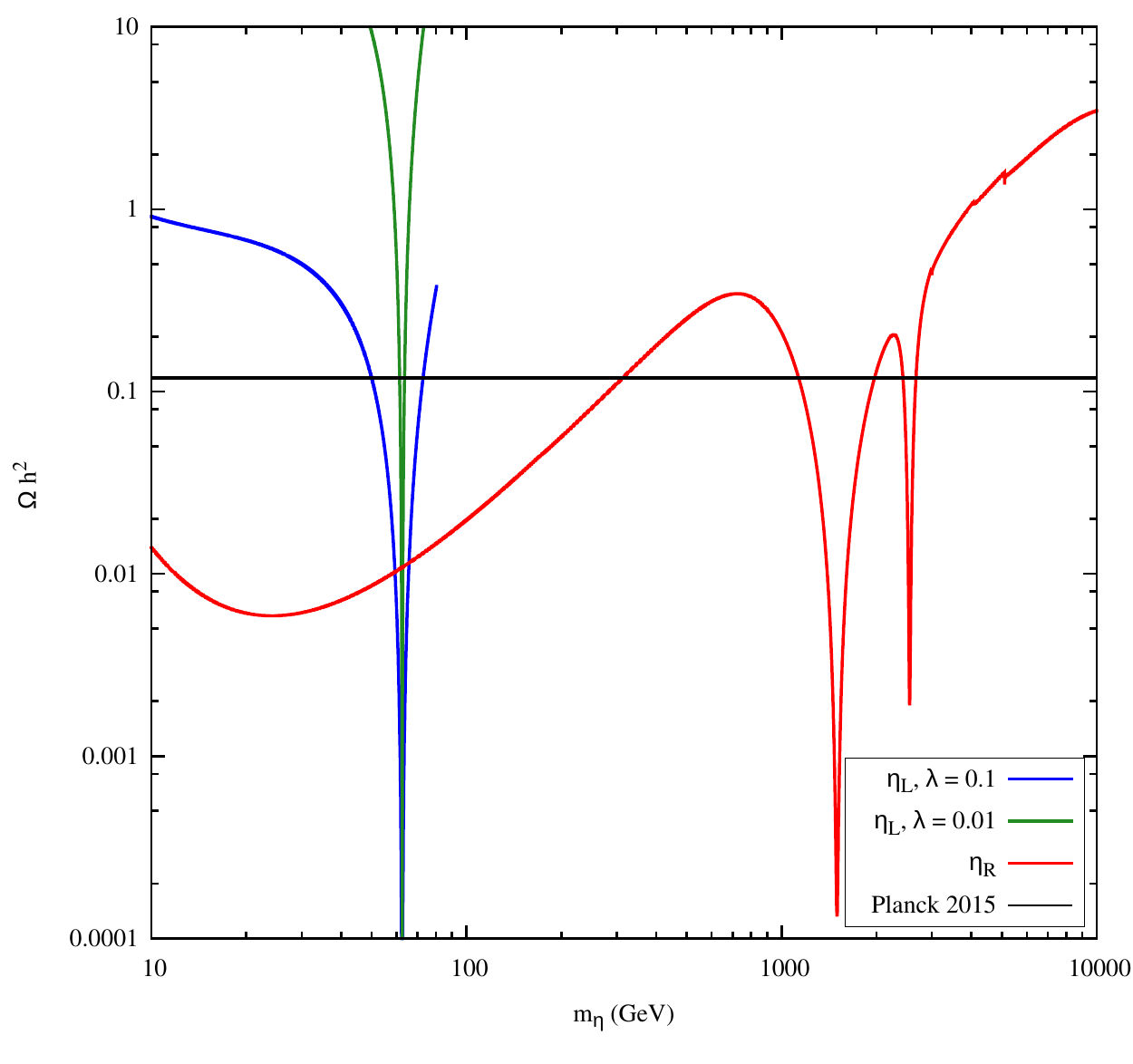}
\caption{Relic abundance of the lightest neutral components of $\eta_L$ and $\eta_R$ scalar doublets. The lightest neutral component of $\eta_L$ is considered to have mass below 80 GeV and annihilating primarily through the Higgs into the SM fermions. The components of $\eta_R$ are assumed to have gauge interactions only, mediated by $W_R, Z_R$ bosons.}
\label{DMrelic}
\end{figure}
\begin{figure}[htb]
\centering
\includegraphics[scale=1.0]{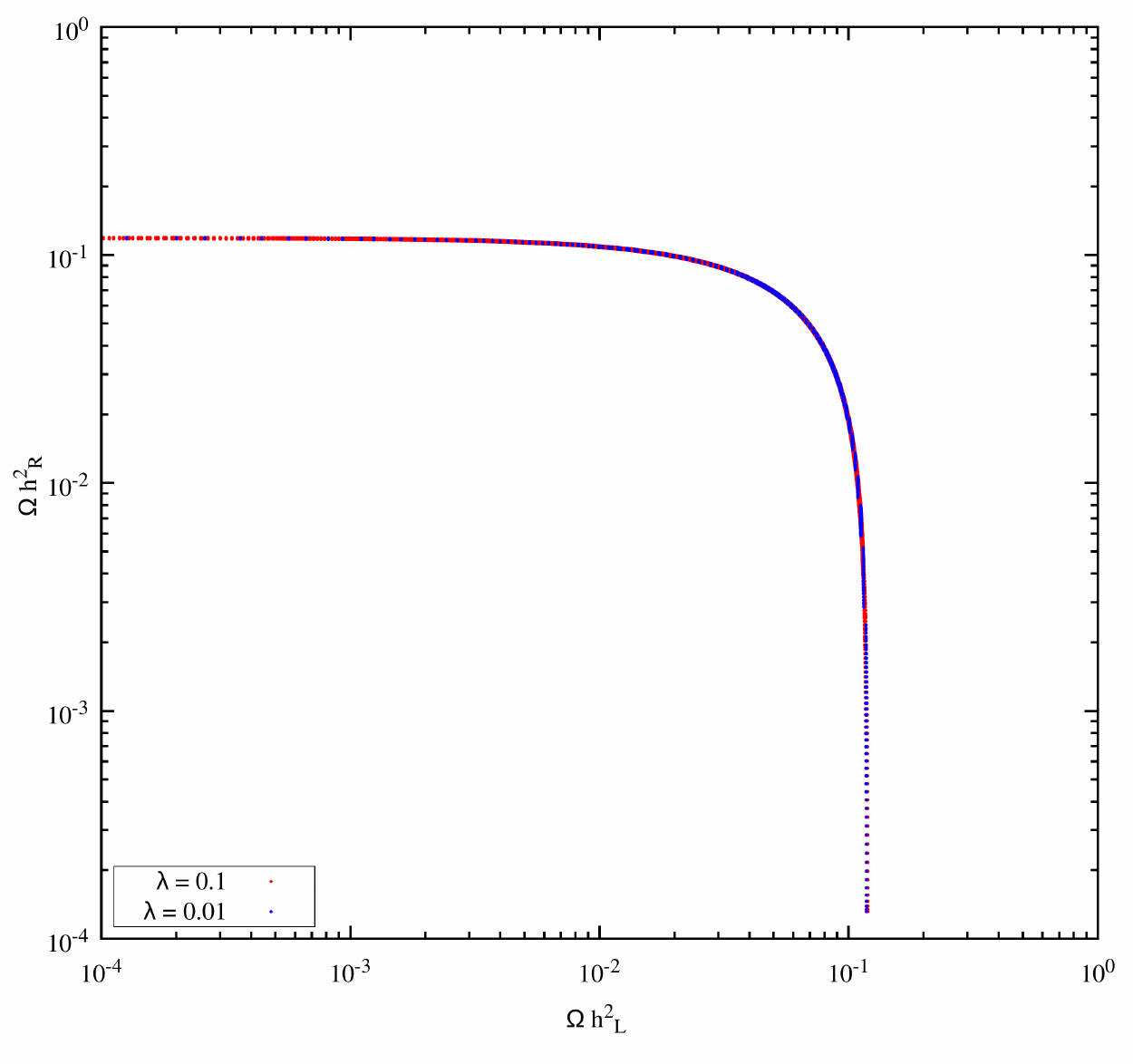}
\caption{Relative contribution of the two dark matter candidates: the lightest neutral components of $\eta_L$ and $\eta_R$ respectively to the total dark matter relic abundance in agreement with the range given by the Planck experiment \eqref{dm_relic}. The left and right scalar dark matter masses are varied in the ranges 10-80 GeV and 10-10000 GeV respectively. }
\label{DMrelic2}
\end{figure}

\begin{figure}[htb]
\centering
\includegraphics[scale=1.0]{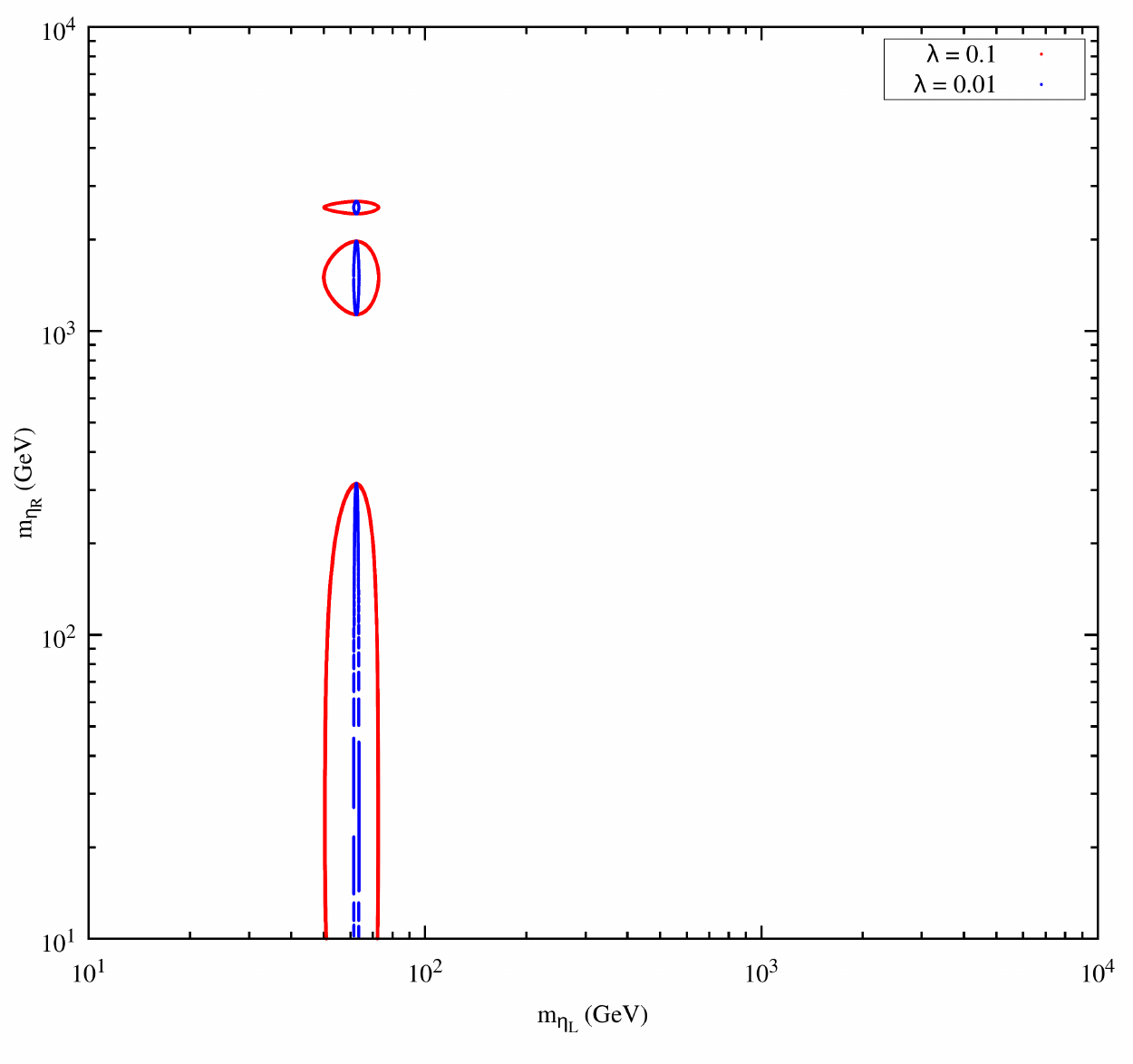}
\caption{Masses of the two dark matter components: the lightest neutral components of $\eta_L$ and $\eta_R$ respectively, when total relic abundance falls within the range given by the Planck experiment \eqref{dm_relic}}
\label{DMrelic3}
\end{figure}

We first show the relic abundance of both $\eta_L$ and $\eta_R$ dark matter as a function of their masses in figure \ref{DMrelic}. We consider both gauge and scalar interactions for $\eta_L$ dark matter. The dominant scalar interactions are the ones through Higgs mediated diagrams and the interaction is parametrised in terms of $\lambda \approx \lambda_{5L}+\lambda_{6L}$. For different values of $\lambda$, the $\eta_L$ contribution to relic abundance changes in the low mass regime $m_{\eta_L} < M_{W_L}$. Above this mass threshold however, the gauge interactions dominate and hence the difference in the DM-Higgs interactions become insignificant, as can be seen from figure \ref{DMrelic}. The resonance region corresponds to $m_{\eta_L} = m_h/2$. The mass splittings between scalar-pseudoscalar as well as charged-neutral scalars are assumed to be high enough so that coannihilations among them are not relevant in case of $\eta_L$ dark matter. For $\eta_R$ dark matter, we consider only gauge interactions and calculate the relic abundance for $M_{W_R} = 3$ TeV. The two different resonance regions correspond to $m_{\eta_R} = M_{W_R}/2, M_{Z_R}/2$ arising due to coannihilations among charged, neutral scalar and neutral pseudoscalar components of $\eta_R$. Our results approximately agree with the ones previously obtained by \cite{Garcia-Cely:2015quu} considering only gauge interactions for both $\eta_L$ and $\eta_R$. We also show the individual contribution of $\eta_L$ and $\eta_R$ to dark matter relic abundance in figure \ref{DMrelic2} such that the total relic abundance agrees with the limit from the Planck experiment \eqref{dm_relic}. The corresponding masses of $\eta_L$ and $\eta_R$ dark matter are shown in figure \ref{DMrelic3} such that the sum of their abundances satisfies the Planck limit.

\begin{figure}[htb]
\centering
\includegraphics[scale=1.0]{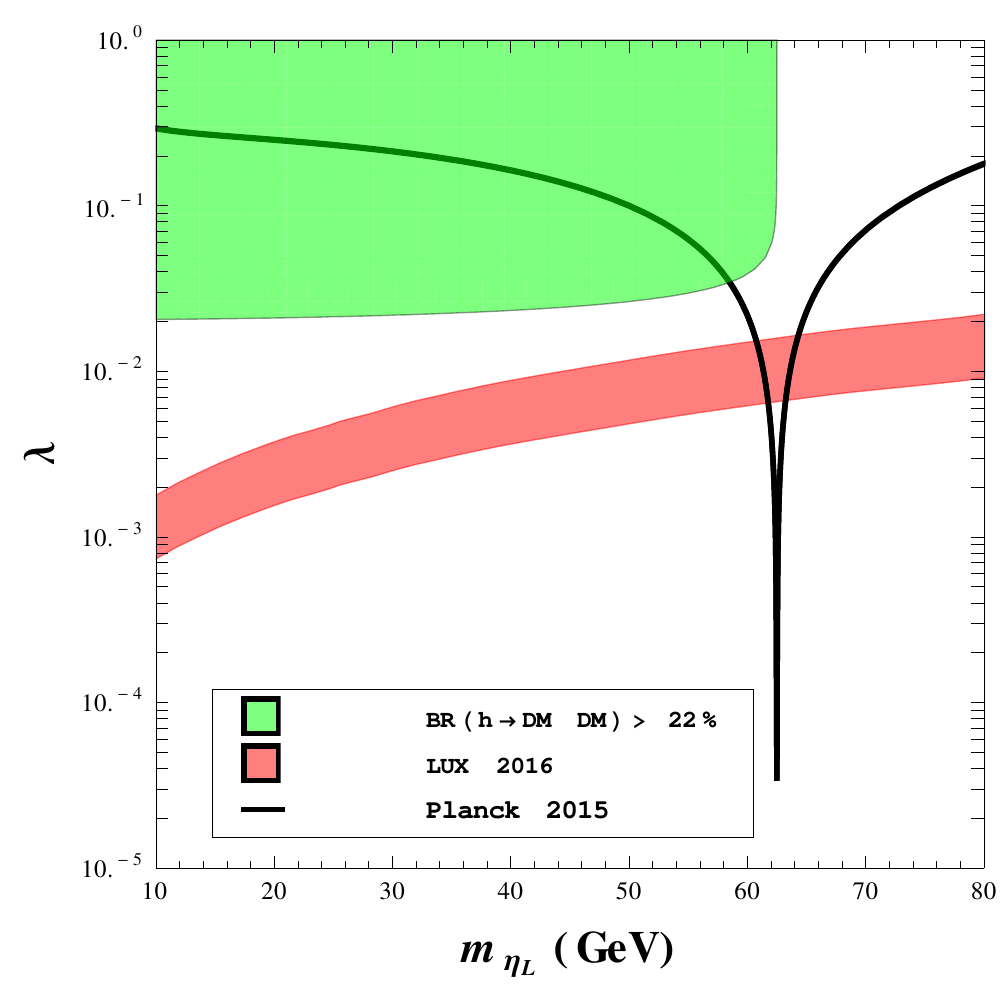}
\caption{Constraint on $\eta_L$-Higgs coupling and $\eta_L$ mass from Planck, LUX and LHC bounds on relic abundance, direct detection cross section and invisible Higgs decay width respectively. $\eta_L$ is assumed to give rise to all the dark matter in the Universe. The thickness of the LUX bound corresponds to the uncertainties in the Higgs-nucleon coupling.}
\label{DMrelic4}
\end{figure}

There also exists bounds from dark matter direct detection experiments like Xenon100 \cite{Aprile:2013doa} and LUX \cite{LUX, LUX16} on the allowed parameter space from relic abundance criteria alone. Since, the right scalar dark matter has only heavy right handed gauge boson interactions and the corresponding mass splitting between different components of the right scalar doublet is assumed to be 1 GeV, there is no tree level dark matter nucleon scattering. However, there can be tree level scattering processes of left scalar dark matter $\eta_L$ with nucleons mediated by the standard model Higgs. The relevant spin independent scattering cross section mediated by SM Higgs is given as \cite{Barbieri:2006dq}
\begin{equation}
 \sigma_{\text{SI}} = \frac{\lambda^2 f^2}{4\pi}\frac{\mu^2 m^2_n}{m^4_h m^2_{\eta_L}}
\label{sigma_dd}
\end{equation}
where $\mu = m_n m_{\eta_L}/(m_n+m_{\eta_L})$ is the $\eta_L$-nucleon reduced mass and $\lambda$ is the quartic coupling involved in $\eta_L$-Higgs interaction which was assumed to take specific values in the relic abundance plot shown in figure \ref{DMrelic}. A recent estimate of the Higgs-nucleon coupling $f$ gives $f = 0.32$ \cite{Giedt:2009mr} although the full range of allowed values is $f=0.26-0.63$ \cite{mambrini}. The latest LUX bound \cite{LUX16} on $\sigma_{\text{SI}}$ constrains the $\eta_L$-Higgs coupling $\lambda$ significantly, if $\eta_L$ gives rise to most of the dark matter in the Universe. According to this latest bound, at a dark matter mass of 50 GeV, dark matter nucleon scattering cross sections above $1.1 \times 10^{-46} \; \text{cm}^2$ are excluded at $90\%$ confidence level. Similar but slightly weaker bound has been reported by the PandaX-II experiment recently \cite{PandaXII}. We however include only the LUX bound in our analysis. One can also constrain the $\eta_L$-Higgs coupling $\lambda$ from the latest LHC constraint on the invisible decay width of the SM Higgs boson. This constraint is applicable only for dark matter mass $m_{\eta_L} < m_h/2$. The invisible decay width is given by
\begin{equation}
\Gamma (h \rightarrow \text{Invisible})= {\lambda^2 v^2\over 64 \pi m_h} 
\sqrt{1-4\,m^2_{\eta_L}/m^2_h}
\end{equation}
The latest ATLAS constraint on invisible Higgs decay is \cite{ATLASinv}
$$\text{BR} (h \rightarrow \text{Invisible}) = \frac{\Gamma (h \rightarrow \text{Invisible})}{\Gamma (h \rightarrow \text{Invisible}) + \Gamma (h \rightarrow \text{SM})} < 22 \%$$
These two constraints on $\eta_L$-Higgs coupling are shown in figure \ref{DMrelic4} where it is assumed that the left scalar dark matter gives rise to all the dark matter in the Universe. The LUX bound incorporated here corresponds to the most conservative one, where we considered the minimum allowed DM-nucleon cross section from \cite{LUX16}. It can be seen that the latest LHC bound is weaker compared to the LUX bound. Incorporating all these experimental constraints makes it clear that, if entire dark matter is in the form of $\eta_L$ and it has mass below $W_L$ mass, then only a small region around $m_h/2 \approx 62.5$ GeV is allowed. This tight constraint on $\eta_L$ mass will become weaker, if $\eta_R$ also contributes substantially to dark matter in the Universe.

\section{Active-Sterile Oscillation}
\label{sec06}
As discussed above, the light neutrinos are predominantly of Dirac type with a tiny Majorana component, leading to the scenario of pseudo-Dirac neutrinos. After integrating out the heavy neutrinos $N_R$, the light neutrino mass matrix in the basis $(\nu_R, \nu_L)$ can be written as
\begin{equation}
\mathcal{M}_{\text{light}}= \left( \begin{array}{cc}
              0 & m_{LR}   \\
              m^T_{LR} & m_{LL}
                      \end{array} \right)
\label{eqn:numatrix2}       
\end{equation}
where $m_{LR} \leq 0.1$ eV is the one-loop Dirac neutrino mass through the Feynman diagram shown in figure \ref{numass2} and $m_{LL} \leq 10^{-13}$ eV is the Majorana mass of left handed neutrinos arising from type I seesaw, whose numerical values are shown in the figure \ref{meevsMajorana}. Since $m_{LR} \gg m_{LL}$, the mass squared difference between two mass eigenstates of the above mass matrix (in one flavour scenario) is $\Delta m^2_{21} \approx 2 m_{LR} m_{LL} \leq 10^{-14} \; \text{eV}^2$. Such tiny pseudo-Dirac splittings can be probed using ultra high energy neutrinos at experiments like IceCube at south pole \cite{activesterile, activesterile1, activesterile2, activesterile3, activesterile4}. However, the usual active neutrino oscillation phenomenology remain unchanged for such tiny mass splitting. For astrophysical neutrinos travelling over large distances like $L\sim 1$ Gpc having energy of the order of PeV, one can probe pseudo-Dirac splitting of the order of $10^{-16}-10^{-15} \; \text{eV}^2$ \cite{activesterile4} which lies in the allowed ranges in our model. The authors of \cite{activesterile4} also pointed out recently that precise future measurement of track-to-shower ratio at next generation IceCube detectors should be able to test such tiny pseudo-Dirac splittings conclusively.

\section{Results and Conclusion}
\label{sec6}
We have studied an extension of the minimal left-right symmetric model where the charged fermions acquire masses through a universal seesaw mechanism, due to the presence of additional heavy vector like fermions. The active neutrinos with the usual $SU(2)_L$ gauge interactions acquire a Dirac mass at one loop level in a scotogenic fashion, such that the lightest among the particles going inside the loop can be a stable dark matter candidate. The particle content of the model augmented by discrete symmetries are chosen in such a way that the active neutrinos form a Dirac fermion $\psi = (\nu_L  \;\; \nu_R)^T$ with $\nu_L$ having $SU(2)_L$ interactions and $\nu_R$ being gauge singlets. The neutral fermion of $SU(2)_R$ lepton doublets however, acquire a heavy Majorana mass from the scalar fields responsible for spontaneous symmetry breaking of LRSM gauge symmetry into the SM one. These heavy neutrino fermions as well as the scalars responsible for their Majorana masses can give rise to observable lepton number violation like neutrinoless double beta decay if the heavy particles are in the TeV region. This non-zero amplitude of $0\nu \beta \beta$ can then generate a tiny Majorana mass of active neutrinos at least at four loop order in accordance with the validity of the Schechter-Valle theorem. We show that, we have a more dominant contribution to the Majorana mass of active neutrinos at two loop order, but that too lies way below the dominant one-loop Dirac mass. Therefore, even for dominantly Dirac nature of active neutrinos, one can realise observable $0\nu \beta \beta$. This scenario is very different from the conventional seesaw models where neutrinos are dominantly Majorana and consequently one can have observable $0\nu \beta \beta$ both from light neutrinos as well as the new physics sector. Although the Schechter-Valle theorem is still valid, this model gives an explicit example showing that the new physics sector responsible for dominant contribution to light neutrino masses and $0\nu \beta \beta$ can be disconnected. Though, the light neutrinos are still Majorana (or pseudo-Dirac), their Majorana masses remain suppressed by several order of magnitudes compared to their Dirac masses. Another complementary probe of dominantly Dirac active neutrinos in the presence of observable $0\nu \beta \beta$ can be provided by cosmology experiments that can distinguish between Dirac and Majorana nature of relic neutrinos \cite{diracNucosmo}.

After discussing the main motivation of the work, we then study the other interesting phenomenology the model provides us with: charged lepton flavour violation and multi-particle dark matter, in particular. We show, how the new physics sector can give rise to observable charged lepton flavour violation like $\mu \rightarrow e \gamma, \mu \rightarrow 3e$. We also show that the present model allows lighter values of triplet scalar mass even after incorporating the latest bounds on LFV decays as well as $0\nu \beta \beta$ half-life. By \textit{lighter values} we mean the values in comparison to previously obtained results. For example, within the minimal LRSM, it was earlier shown that \cite{ndbd00} the triplet scalar mass should be at least ten times heavier than the heaviest neutral lepton. This was subsequently shown to be at least two times \cite{ndbd102} and even equal \cite{ndbd103}. Here, we have shown that the scalar triplet can even be ten times lighter than the heaviest neutral lepton. We finally consider the interesting dark matter sector in the model, which simultaneously allow one left and one right scalar doublets to be stable dark matter candidates, a feature which is not there in the minimal LRSM augmented by two scalar doublets. For simplicity, we consider negligible scalar couplings between the two sectors and also neglect the scalar coupling contribution to right handed scalar dark matter. By considering the interactions of $\eta_L$ dark matter with SM Higgs and electroweak gauge bosons, we calculate the relic abundance and show two different region of masses where it can give rise to the total relic abundance. For $\eta_R$ dark matter, we consider only the heavy right handed gauge boson interactions and calculate its relic abundance for $M_{W_R} = 3$ TeV. We also show their individual contributions to total dark matter abundance such that the total relic abundance agrees with observations. The corresponding values of their masses are also shown. We find that, even for such simplistic assumptions of couplings, we get a wide region of parameter space that can give rise to the observed relic abundance. Allowing any sizeable interactions between left and right sector dark matter candidates should open up more region of parameter space. Since this involves a complicated calculation of coupled Boltzmann equations for the two dark matter candidates, we leave this detailed study for a future work. Such multi-particle dark matter can also give rise to interesting collider phenomenology, as their individual production cross sections can be significantly enhanced compared to single component dark matter scenarios. Another interesting future direction could be the study of the origin of matter-antimatter asymmetry within such frameworks. Since, the light neutrinos are predominantly Dirac, one can perhaps consider the possibility of generating matter antimatter asymmetry of the Universe through Dirac leptogenesis \cite{diraclepto0,dbad1}. These interesting possibilities are left for a future work.

\begin{acknowledgments}
DB would like to express a special thanks to the Mainz Institute for Theoretical Physics (MITP) for its hospitality and support during the workshop \textit{Exploring the Energy Ladder of the Universe} where this work was initiated. DB also thanks Alexander Merle for very useful discussions about the Schechter-Valle theorem and Julian Heeck for discussions about the calculation of dark matter relic abundance.
\end{acknowledgments}
\bibliographystyle{apsrev}

\end{document}